\documentclass[aps, prl, reprint,10pt,longbibliography,superscriptaddress]{revtex4-2}

\usepackage[caption=false]{subfig}

\usepackage{amsmath, amssymb, amsfonts, amsthm, bm, mathrsfs, bbm, enumerate}
\usepackage{graphicx}
\usepackage{epstopdf}
\usepackage{color}
\usepackage[usenames,dvipsnames]{xcolor}
\usepackage[citecolor=blue, colorlinks=true, urlcolor=blue, linkcolor=blue]{hyperref}
\usepackage{bbold}
\usepackage{feynmp-auto} 

\def \ETH{Institute for Quantum Electronics, ETH Z\"urich, CH-8093 Z\"urich, Switzerland}

\begin{document}
\title{Tunable Feshbach resonances and their spectral signatures in bilayer semiconductors}

\author{Clemens Kuhlenkamp}
\affiliation{\ETH}
\affiliation{Department of Physics and Institute for Advanced Study, Technical University of Munich, 85748 Garching, Germany}
\affiliation{Munich Center for Quantum Science and Technology (MCQST), Schellingstr. 4, D-80799 M{\"u}nchen, Germany}
\author{Michael Knap}
\affiliation{Department of Physics and Institute for Advanced Study, Technical University of Munich, 85748 Garching, Germany}
\affiliation{Munich Center for Quantum Science and Technology (MCQST), Schellingstr. 4, D-80799 M{\"u}nchen, Germany}
\author{Marcel Wagner}
\affiliation{Max-Planck-Institute of Quantum Optics, Hans-Kopfermann-Straße 1, 85748 Garching, Germany}
\affiliation{Munich Center for Quantum Science and Technology (MCQST), Schellingstr. 4, D-80799 M{\"u}nchen, Germany}
\author{Richard Schmidt}
\affiliation{Max-Planck-Institute of Quantum Optics, Hans-Kopfermann-Straße 1, 85748 Garching, Germany}
\affiliation{Munich Center for Quantum Science and Technology (MCQST), Schellingstr. 4, D-80799 M{\"u}nchen, Germany}
\author{Ata\c{c} Imamo\u{g}lu}
\affiliation{\ETH}

\begin{abstract}
Feshbach resonances are an invaluable tool in atomic physics, enabling precise control of interactions and the preparation of complex quantum phases of matter. Here, we theoretically analyze a solid-state analogue of a Feshbach resonance in two dimensional semiconductor heterostructures. In the presence of inter-layer electron tunneling, the scattering of excitons and electrons occupying different layers can be resonantly enhanced by tuning an applied electric field. The emergence of an inter-layer Feshbach molecule modifies the optical excitation spectrum, and can be understood in terms of Fermi polaron formation. We discuss potential implications for the realization of correlated Bose-Fermi mixtures in bilayer semiconductors.
\end{abstract}

\date{\today}

\pacs{
}

\maketitle

Recently bilayer structures of two-dimensional materials have emerged as fascinating platforms for realizing exotic phases of electronic matter~\cite{moire_rev,liu20_bilayer}. Much of their success is driven by a new level of control, arising from twisting the two layers with respect to each other during stacking. Such twisted bilayers generate a moir\'e potential for electrons or holes, which quenches the kinetic energy and therefore enhances correlations. Most notably this has lead to the discovery of unconventional superconductivity~\cite{cao_18,bistritzer11}, correlated insulators, and charge density waves~\cite{regan_20,tang_20,yuya_20} in bilayer graphene and transition metal dichalcogenides (TMDs). 
In addition to electronic phases, semiconductors such as TMDs can host excitons, which are strongly bound electron-hole pairs. They act as mobile composite bosons and remain rigid due to their large binding energies. Moreover, excitons interact with free electrons or holes and can form charged molecules, termed trions. This renders bilayer TMDs promising candidates to study complex Bose-Fermi mixtures. Such mixtures have been recently investigated in dilute quantum gases~\cite{barbut15,desalvo19,fritsche21}, where Feshbach resonances are routinely used to control interactions between the atomic species~\cite{feshbach1962,chin_rev_10,ketterle_08,zwerger_rev_08,duine_04}. By contrast, in solid state structures the molecular binding energies, and correspondingly the interaction strength among particles, are generically fixed by material properties, limiting the experimentally accessible regimes. 

Here, we address this challenge by introducing a solid-state analogue of a Feshbach resonance. Using the layer degree of freedom as a pseudo-spin, we demonstrate that the energy of a closed-channel bound state can be tuned with respect to scattering states in an open channel, simply by applying an external electric field $E_z$. The counterpart of hyperfine interactions in atomic systems, is provided by coherent inter-layer electron or hole tunneling. We demonstrate the impact of such Feshbach resonances on the spectrum of a single optically-injected exciton immersed in a Fermi sea of charge carriers, taking into account the radiative exciton decay. Close to the Feshbach resonance we find a striking modification of the exciton spectrum. In particular, we show that the spectral shape is sensitive to the finite range of the effective interactions relative to the Fermi wavelength.

Our work is motivated by a recent experimental observation of an electrically tunable Feshbach resonance in a twisted bilayer TMD heterostructure~\cite{schwartz21}. We theoretically analyze a more generic scenario with vanishing twist angles and discuss how resonantly enhanced polaron formation can be observed in reflection measurements. Our findings demonstrate the potential for bilayer TMDs to control valley-selective interactions between itinerant carriers and establish a novel platform for exploring correlated quantum dynamics of degenerate Bose-Fermi mixtures.

\begin{figure}[t]
\includegraphics[width=1\columnwidth]{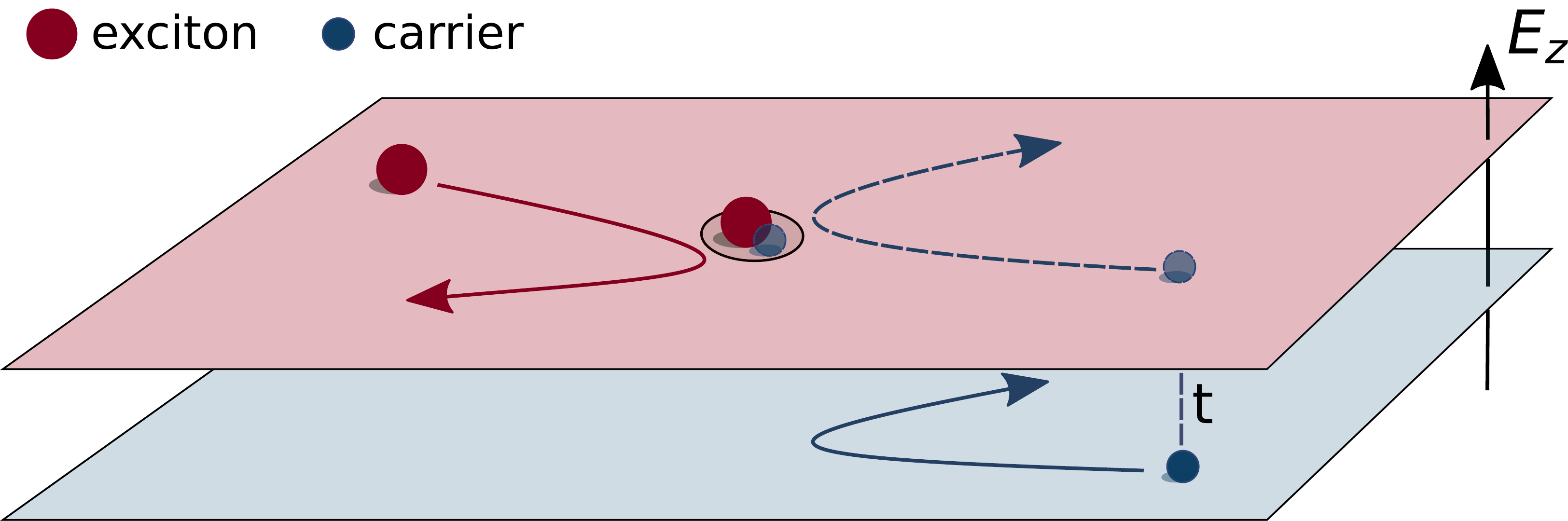}	\caption{\textbf{Tunable Feshbach resonances in bilayer heterostructures.} Illustration of exciton-carrier scattering in a bilayer TMD. The electrostatic potential energy is different in the two layers and can be tuned by a perpendicular electric field $E_z$. Scattering between excitons and electrons is enhanced when the intra-layer trion energy is tuned into resonance with the energy of an electron and an exciton in seperate layers.}
\label{fig:1}
\end{figure}

\textbf{Effective bilayer Hamiltonian.---} We consider a bilayer semiconductor setup as depicted in Fig.~\ref{fig:1}. Electrons (or holes) can tunnel from one layer to the other with coupling constant $t$, which can be adjusted by incorporating tunnel barriers~\cite{yuya_20}. For concreteness we focus on TMD bilayers spatially separated by a distance $d$. Generically, the exciton resonances in the top and bottom layers have different energies, either due to difference in material properties or strain, enabling layer-selective exciton creation. Furthermore, hybridization between inter- and intra-layer excitons is small, due to the sizable difference in binding energies. This allows us to focus only on intra-layer excitons~\cite{Note4}.\footnotetext[4]{Large electric fields could overcome the energy difference resulting in sizable hybridization with inter-layer excitons, which would then contribute to inter-layer scattering of electrons and excitons. Here are we are interested in smaller fields on the order of the trion binding energy where these processes are suppressed.} For simplicity we assume that excitons are injected optically and are present only in the top layer. The system is then described by the effective Hamiltonian
\begin{equation}
\begin{aligned}
\hat{H} =& \sum_\mathbf{k} x^\dagger_\mathbf{k} \frac{k^2}{2M} x_\mathbf{k} +\begin{pmatrix}c^\dagger_{\mathbf{k},T} \\ 
c^\dagger_{\mathbf{k},B}\end{pmatrix} \begin{pmatrix} \xi_\mathbf{k}+ \Delta & t  \\ t & \xi_\mathbf{k}\end{pmatrix}\begin{pmatrix} c_{\mathbf{k},T} \\ 
c_{\mathbf{k},B}\end{pmatrix}\\
&+ \frac{U}{V} \sum_{kk'q} c_{\mathbf{k},T}^\dagger c_{\mathbf{k}+\mathbf{q},T} x^\dagger_{\mathbf{k'}} x_{\mathbf{k'}-\mathbf{q}},
\label{Eq:BoseFermiHamiltonian}
\end{aligned}
\end{equation}
where $x_\mathbf{k}^\dagger$ creates an exciton of mass $M$ in the top layer, and $c^\dagger_{\mathbf{k},T}$ and $c^\dagger_{\mathbf{k},B} $ create fermions of mass $m$ in the top and bottom layer, respectively. From now on we refer to itinerant charges as electrons, although all conclusions apply equally to holes. We omit the valley and spin degree of freedom and assume that electrons and excitons reside in different valleys, since only this scattering channel will be resonantly enhanced.
As the exciton’s Bohr radius is small, excitons and electrons experience sizable attractive contact interactions $U$, only when both particles are in the same layer and opposite valleys.  We also neglect the composite nature of the exciton and treat it as a structureless boson~\cite{Fey20}. The potential energy difference $\Delta = q d E_z$ between the two layers, can be tuned by changing $E_z$, as illustrated in Fig.~\ref{fig:1}. We consider the scenario where  $\Delta$ is chosen such that electrons reside predominantly in the bottom layer.

\begin{figure}
\includegraphics[width=1\columnwidth]{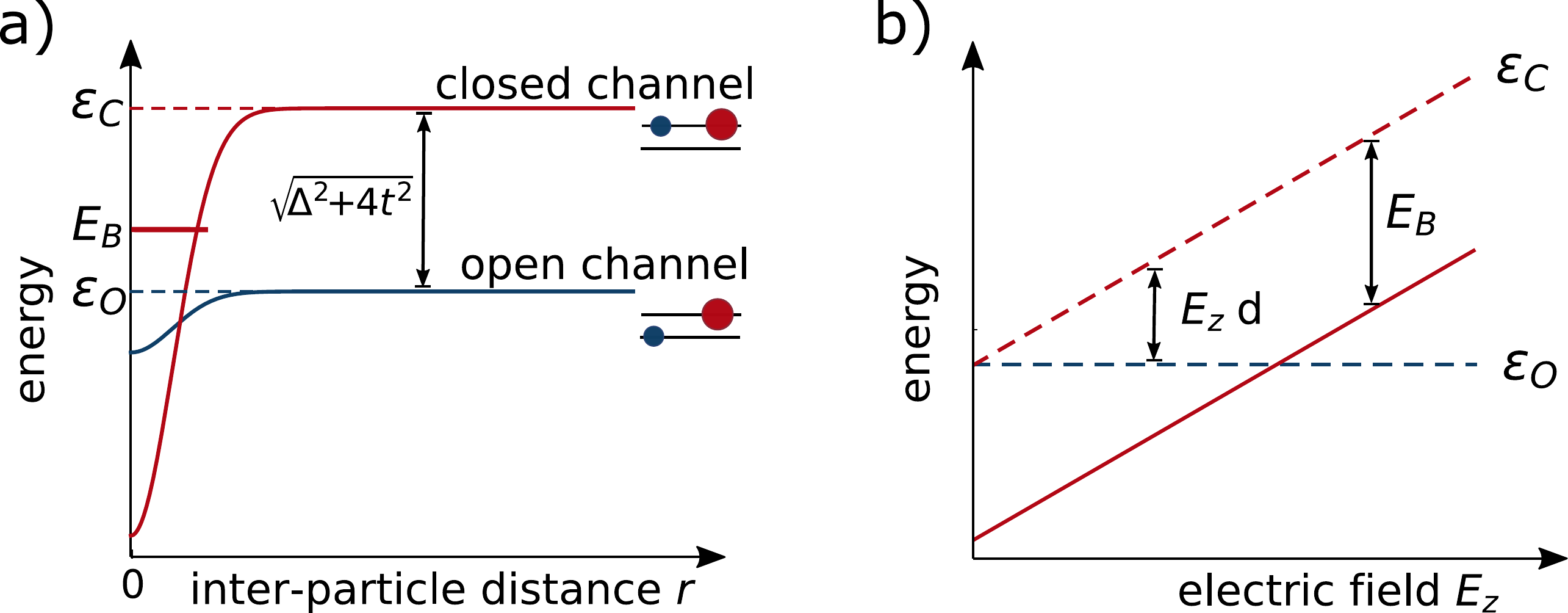}
\caption{\textbf{Illustration of scattering channels.} \textbf{(a)} Inter-particle potential for an exciton and an electron prepared in the open (blue) or closed (red) channel.   \textbf{(b)} Threshold energies of the open and closed channel $\varepsilon_O$ and $\varepsilon_C$, as the electric field is varied. The bare closed channel bound-state energy is denoted as a red dashed line. This bound state can be brought into resonance with $\varepsilon_O$ for an appropriately chosen electric field.}
\label{fig:2}
\end{figure}

\textbf{Feshbach resonance in exciton-electron scattering.---} To understand scattering properties in such a heterostructure, we focus on the two-particle subspace of the system. In the center of mass frame, Eq.~\ref{Eq:BoseFermiHamiltonian} can then be expressed in first quantization as:
\begin{equation}
\begin{aligned}
&\hat{H}_{\mathrm{2\,body}} =\hat{H}_0 + \hat{U}\\
&= \begin{pmatrix} -\frac{\mathbf{\nabla}_{\mathbf{R}}^2}{2m_{\mathrm{tot}}} -\frac{\mathbf{\nabla}_{\mathbf{r}}^2}{2\mu} +\Delta & t  \\t &-\frac{\mathbf{\nabla}_{\mathbf{R}}^2}{2m_{\mathrm{tot}}} -\frac{\mathbf{\nabla}_{\mathbf{r}}^2}{2\mu}\end{pmatrix} +U\begin{pmatrix} \delta^2(\mathbf{r})& 0  \\ 0 &0
\end{pmatrix},
\end{aligned}
    \label{Eq:feshbach_firstq}
\end{equation}
where $\mu = 1/(m^{-1} + M^{-1})$ and $m_{\mathrm{tot}}= m+M$ are the reduced mass and the total mass respectively. The wave function carries the layer degree of freedom and the part describing the relative motion can be expressed as $\psi(\mathbf{r}) = (\psi_T(\mathbf{r}),\psi_B(\mathbf{r}))^T/\sqrt{2}$. Asymptotic eigenstates with large spatial separation between the two particles define the open and closed channel. We consider $E_z$ for which $\Delta\simeq |E_B^0|$, where $E_B^0$ is the binding energy of the intra-layer trion. Although both channels are hybridized between the layers, only the open channel is energetically accessible and electrons reside predominantly in the bottom layer (Fig.~\ref{fig:2}). The scattering threshold for the open ($\varepsilon_O$) and closed ($\varepsilon_C$) channel is $\varepsilon_{O,C} = \Delta/2  \mp \sqrt{t^2 +\Delta^2/4} $.

The outgoing scattering states $|\psi^+_\alpha\rangle$, in channel $\alpha$ with energy $E$ can be found as solutions of the Lippmann-Schwinger equation:
\begin{equation}
|\psi^+_\alpha\rangle = |\phi_\alpha\rangle +  \frac{1}{E-\hat{H}_0 + i0^+ } \hat{U} |\psi^+_\alpha\rangle,
\label{Eq:lippmann_schwinger}
\end{equation}
where $\langle \mathbf{r}|\phi_\alpha \rangle \sim e^{i k x} $ is an incoming plane wave. We can reformulate the problem by introducing the T-matrix $\hat{T}^R|\phi_\alpha\rangle = \hat{U} |\psi_\alpha^+\rangle$, which connects the incoming plane waves with the full outgoing scattering state. Eq.~\ref{Eq:lippmann_schwinger} translates to an equation for the off-shell T-matrix $\hat{T}^R(E)$:
\begin{equation}
\hat{T}^R(E) = \hat{U} + \hat{U} \left(E - \hat{H}_0 + i 0^+ \right)^{-1} \hat{T}^R(E).
    \label{Eq:tmatrix}
\end{equation}
 We solve Eq.~\ref{Eq:tmatrix} analytically in a plane-wave basis which diagonalizes $\hat{H}_0$: 
\begin{equation}
    \begin{aligned}
    \hat{T}^R(E,\mathbf{k}) &= [\mathbb{1}_{2\times2}
 - \hat{U} \cdot \Pi^R(E, \mathbf{k})]^{-1} \cdot \hat{U} \\
 \Pi^R_{\alpha\beta}(E, \mathbf{k}) &=\int \frac{d^2q}{(2\pi)^2} \frac{\delta_{\alpha\beta}}{E -\frac{\mathbf{q}^2}{2\mu}- \frac{\mathbf{k}^2}{2m_{\mathrm{tot}}}-\varepsilon_\alpha  + i 0^+},
    \end{aligned}
\label{Eq:tmatrix_2p}
\end{equation} 
where $E$ is the scattering energy, and $\mathbf{k}$ is the total incoming momentum. The $2\times2$ matrix structure of $\hat{T}^R(E,\mathbf{k})$, 
and $\hat{U}$,
due to the two channels, is implicitly assumed.

\begin{figure}[t]
\includegraphics[width=1\columnwidth]{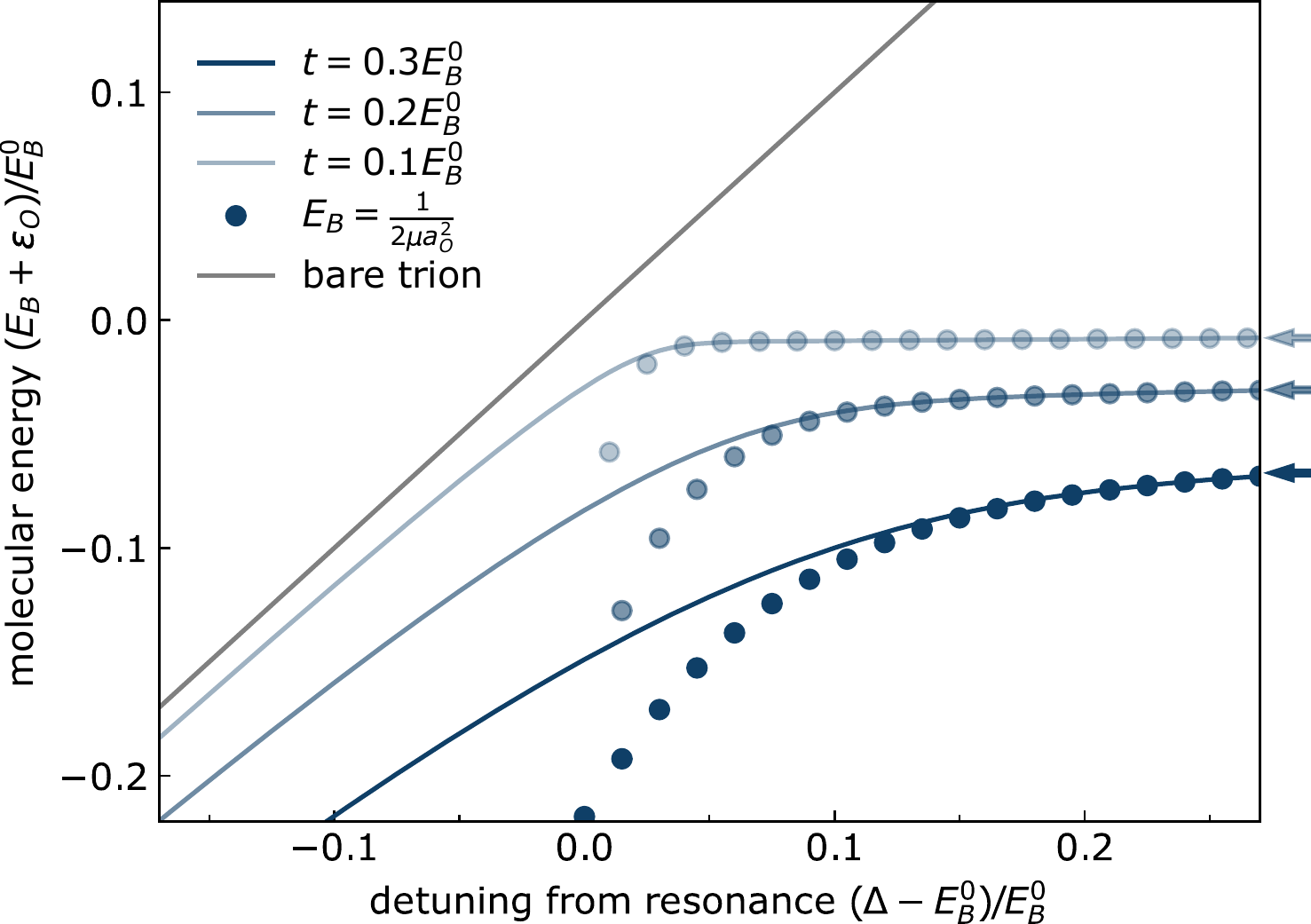}
\caption{\textbf{Feshbach molecule binding energy.} Molecular energy as a function of electric field (solid blue lines). We have assumed an exciton mass of $M = 2m$ and contact interactions between the exciton and electron. In two dimensions, and in the absence of inter-layer repulsion, a bound state exists for all values of $\Delta$. When the size of the molecule exceeds the the range of the interactions, the scattering length alone determines the binding energy (blue dots). For large positive detunings the molecular energy approaches the open channel threshold $\varepsilon_O$ (arrows), implying that the binding energy $E_B$ goes asymptotically to zero. For large negative detunings the binding energy approaches the energy of the bare intra-layer trion.}
\label{fig:3}
\end{figure}

Scattering can be resonantly enhanced if $E_z$ is tuned such that the closed channel bound state is in proximity of the open channel threshold $\varepsilon_O$, see Fig.~\ref{fig:2} (b) for an illustration. Similar to cold atomic systems, we are interested in two-particle collisions with small incoming momenta. In this case, scattering is accurately described by a finite-range expansion, which is performed by expanding the denominator of the T-matrix in powers of $E-\varepsilon_O$. In two dimensions the finite range expansion of the on-shell T-matrix takes the universal form:
\begin{equation}
T^R(\mathbf{k}^2/2\mu,\mathbf{0})^{-1} =  \frac{\mu}{2\pi} \left( i\pi - \ln(\mathbf{k}^2 a^2) + \frac{r_0 \mathbf{k}^2}{2} + \mathcal{O}(\mathbf{k}^3)  \right),
    \label{Eq:finiterangeexp}
\end{equation}
which is characterized by the scattering length $a$ and effective range $r_0$~\cite{adhikari}. We relate this expansion to our effective description by integrating Eq.~\ref{Eq:tmatrix_2p} and matching the open channel scattering amplitude $T^R_{OO}(\mathbf{k}^2/2\mu,\mathbf{0})$ to Eq.~\ref{Eq:finiterangeexp}. This way we obtain the open channel scattering length $a_O$ and effective range $r_0$:
\begin{equation}
\begin{aligned}
a_O &= a \exp\left\lbrace -\frac{1}{2} \left(\frac{\Delta}{2t} + \sqrt{1+\frac{\Delta^2}{4t^2}}\right)^2 \ln \frac{-E_B^0}{\sqrt{4t^2 + \Delta^2}}\right\rbrace, \\
r_0 &= \frac{1}{2\mu} \frac{(\Delta/2 + \sqrt{t^2 + \Delta^2/4})^2}{t^2\sqrt{t^2 + \Delta^2/4}},
\end{aligned}
\label{Eq:scatteringlength}
\end{equation}
where $a = \sqrt{2\mu E_B^0}$ is the scattering length of the closed channel in absence of tunnel coupling. Analyzing Eq.~\ref{Eq:tmatrix_2p}--\ref{Eq:scatteringlength}
we find that the open channel T-matrix has a pole at energies below the scattering threshold $\varepsilon_O$. This is the signature of a Feshbach molecule which forms in inter-layer scattering~\cite{Weinberg}.  Eq.~\ref{Eq:finiterangeexp} demonstrates that the energy of the molecule depends on both the scattering length $a_O$ and range $r_0$. We plot the energy of the Feshbach molecule as a function of detuning in Fig.~\ref{fig:3} for three different $t$. As the detuning becomes large and positive, the scattering length starts to diverge while the molecular energy approaches the scattering threshold. For large detunings the binding energy is then approximately given by $1/2\mu a_O^2$. In the case $E_B^0, \Delta \gg t$ we obtain simple expressions for the binding energy of the Feshbach molecule and the effective range close to resonance, which read
\begin{equation}
\begin{aligned}
E_B \simeq E_B^0 \frac{1}{e^{-2}} \left|\frac{\Delta}{E_B^0}\right|^{-\frac{\Delta^2}{t^2}},\qquad
r_0 \simeq \frac{1}{ \mu} \frac{\Delta}{t^2}.
\end{aligned}
    \label{Eq:approx_binding}
\end{equation}
This demonstrates the power of a Feshbach resonance: complete control over the energy of the Feshbach molecule can be achieved  simply by changing $E_z$. Thus the system can be electrically tuned to arbitrarily large scattering lengths~\cite{Note1,kanjilal06}. \footnotetext[1]{More importantly, for any given electron density, it is possible to choose $E_z$ such that $k_F a_O \sim 1$, which yields the strong correlation regime.} While the binding energy of the Feshbach molecule changes exponentially, the effective range $r_0$ depends only linearly on $E_z$. We find that weakly coupled layers lead to large values of $r_0$, and the resulting physics is reminiscent of narrow Feshbach resonances in three dimensions.

\begin{figure*}[t] 
\includegraphics[width=2\columnwidth]{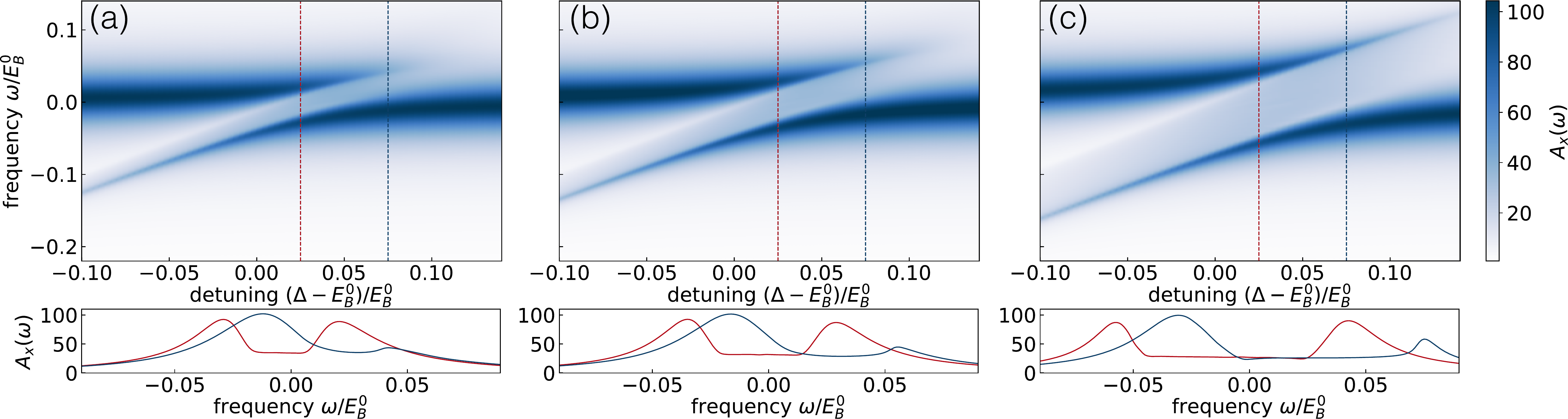}
\caption{\textbf{Exciton spectra across the Feshbach resonance.} The spectral function of a dissipative exciton as a function of the bias $\Delta$, computed within a T-matrix approximation. The Fermi energy $E_F$ is increasing from left to right: \textbf{(a)} $E_F = E_B^0/30$, \textbf{(b)} $E_F = E_B^0/20$, \textbf{(c)} $E_F = E_B^0/10$. All spectra are computed for weak channel coupling $t = 0.15 E_B^0$. The splitting of the repulsive and attractive branch depends on $E_F$, as highlighted in the line cuts of the spectra for two different $\Delta$ in the lower panels. For large $E_F$, finite range corrections become increasingly important and the spectral shape starts to resemble an anticrossing. The exciton is assumed to have a radiative lifetime of $\Gamma= E_B^0/30$.}
\label{fig:4}
\end{figure*}
In contrast to the three dimensional case however, the bound state does not dissolve for any value of $E_z$, as long as there is no repulsive background scattering. The purely two dimensional geometry of the system also distinguishes the proposed resonance from realizations in cold atom systems, where scattering always remains effectively three dimensional due to the finite transverse confinement~\cite{zwerger_rev_08}.

\textbf{Optical impurities strongly coupled to a Fermi sea.---}
Resonantly enhanced two-particle scattering affects correlations in electron-exciton mixtures. We consider a low concentration of excitons injected into a Fermi sea of electrons in the open channel. The excitons in such a system are mobile impurities and form collective excitations known as Fermi polarons~\cite{sidler_polaron,suris_03, efimkin}. Here we analyze the polaron spectrum as $E_z$ is tuned over the Feshbach resonance.

Our previous discussions focused on two-body scattering with small but finite momentum, for which the exciton is a long-lived excitation and the scattering matrix is essentially unitary~\cite{Note2}.\footnotetext[2]{Finite momentum excitons are long-lived due to the steep light-cone, which renders only $\mathbf{k}=0$ excitons optically active.} Here, we focus on optically excited $\mathbf{k}=0$ excitons. In this regime excitons couple to the radiation field, which allows them to decay via electron-hole recombination via emission of an optical photon. As this decay process is essentially memory-less, it can be described by a Lindblad master equation
\begin{equation}
\begin{aligned}
\dot{\rho}(t)&= -i [\hat{H},\rho] + \sum_{\mathbf{k}} L_\mathbf{k} \rho L_\mathbf{k}^\dagger - \frac{1}{2}\lbrace L_\mathbf{k}^\dagger L_\mathbf{k},\rho \rbrace, \\
L_\mathbf{k} &= \sqrt{2\Gamma(\mathbf{k})}\,x_\mathbf{k}\;,
\end{aligned}
\label{Eq:lindblad}
\end{equation}
where $\Gamma(\mathbf{k})$ is the decay rate of the exciton. In the presence of a Fermi sea Eq.~\ref{Eq:lindblad} constitutes a complex many-body system, which can not be solved exactly. However, it was found that key properties can already be inferred purely from the scattering properties of the system~\cite{fumi_thm,schmidt_rev_2018} and that T-matrix approximations provide an accurate description of the ground and excited states of mobile impurities~\cite{chevy,combescot,schmidt_polaron2d,qmc_1,qmc_2,cetina_16,Parish2011,Parish2013,Zoellner2011}.

For our heterostructure setting we develop a T-matrix approximation to include dissipation as well as finite-range corrections from the Feshbach resonance:
\begin{equation}
\begin{aligned}
 \hat{T}^R(E,\mathbf{k}) &= [\mathbb{1}_{2\times2}
 - \hat{U} \cdot \Pi^R(E, \mathbf{k})]^{-1} \cdot \hat{U} \\
 \Pi^R_{\alpha\beta}(E, \mathbf{k}) &=\int\limits_{|\mathbf{l}|>k_F} \frac{d^2l}{(2\pi)^2} \frac{\delta_{\alpha\beta}}{E -\xi_{\mathbf{l}} -\varepsilon_\alpha- \frac{(\mathbf{k}-\mathbf{l})^2}{2M} + i \Gamma(\mathbf{k}-\mathbf{l})}.
\label{Eq:in_medium_tmatrix}
\end{aligned}
\end{equation}
Details on the calculation can be found in the supplemental materials~\cite{supp}. Compared to Eq.~\ref{Eq:tmatrix_2p}, the momentum of the electron in the open channel is now restricted to lie above the Fermi surface due to Pauli blocking by the Fermi sea. Exciton recombination results in an imaginary part $i \Gamma(\mathbf{k})$ of the exciton energy~\cite{sieberer_16,wasak_21}. Using this T-matrix, we then determine the self-energy of the exciton as a function of frequency $\omega$:
\begin{equation}
\Sigma^R(\omega,\mathbf{k}) = \int\limits_{|\mathbf{q}|<k_F} \frac{d^2 q}{(2\pi)^2} T^R_{OO}(\omega + \xi_\mathbf{q}, \mathbf{k}+\mathbf{q}).
\end{equation} This equation originates from the creation of a particle-hole pair in the open channel, with hole momentum $\mathbf{q}<k_F$. The spectral function of the exciton then reads
\begin{equation}
A_x(\omega,\mathbf{k}) = \mathrm{Im}\left[ \frac{-\pi}{\omega - \mathbf{k}^2/2M - \Sigma^R(\omega,\mathbf{k}) + i \Gamma(\mathbf{k})}\right].
\label{Eq:spectral_function}
\end{equation}
 As the master equation fulfills fluctuation-dissipation relations and we have treated dissipation exactly, the resulting spectral function respects the sum rule $\int \frac{d \omega}{2\pi}\,A_x(\omega,\mathbf{k}) = \langle [x_\mathbf{k},x_\mathbf{k}^\dagger] \rangle= 1 $.

We compute the spectrum as a function of detuning, by integrating Eq.~\ref{Eq:spectral_function} numerically. We show the resulting exciton spectra in Fig.~\ref{fig:4} for three different Fermi energies (a)-(c). They are characterized by the formation of an attractive branch, with maxima at negative frequencies; and a repulsive branch, with maxima at positive frequencies. For small Fermi energies (Fig.~\ref{fig:4}(a)) the two resonances approach the Feshbach molecule and bare exciton energy respectively: the spectrum can be understood in terms of the formation of a Fermi polaron and is clearly distinct from an avoided level crossing. We observe that the repulsive polaron abruptly loses spectral weight and blue-shifts in energy, as the Feshbach molecule becomes weakly bound. With increasing carrier density, the maximal splitting between the repulsive and attractive branch  grows (Fig.~\ref{fig:4}(b)). Surprisingly, the spectrum at high densities starts to resemble an avoided crossing (see Fig.~\ref{fig:4}(c)). This change in spectral shape cannot be explained assuming contact interactions, but rather arises from significant finite range corrections~\cite{kohstall12,massignan12}. Since the average scattering process involves momenta on the order of $k_F$, the non-logarithmic terms in Eq.~\ref{Eq:finiterangeexp} become successively more important at high densities and strongly renormalize the spectrum. The spectral asymmetry we observe is distinct from solid-state realizations based on polaritons, whose steep dispersion obscures the relevant scattering physics~\cite{takemura_14}. As the spectral function of the exciton is directly accessible in reflection measurements, the features we identified provide particularly clear experimental signatures, which result from many-body effects. 

\textbf{Conclusions and Outlook.---} 
We have investigated an electrically tunable solid-state Feshbach resonance, using the layer pseudo-spin degree of freedom of semiconductor bilayers. Our scheme allows for a controlled enhancement of electron-exciton scattering in experiments. We find that much of the resulting Feshbach physics, such as Fermi-Polaron formation for a dilute concentration of excitons, may be observed in experiments as a highly asymmetric and density-dependent reflection spectrum. This makes TMD bilayers ideal systems to study two dimensional Fermi polarons in parameter regimes that have so far been inaccessible in cold atomic gases and monolayer semiconductors. By extending our setup to finite exciton densities, Feshbach resonances could enable precise control of degenerate Bose-Fermi mixtures in solid state systems. This is particularly appealing as excitations of an excitonic Bose gas can mediate superconductivity in a Fermi sea~\cite{kavokin_10,ovidiu_sc_16}. Since the bound state exists only for excitons and electrons with different spin$/$valley degree-of-freedom, the Feshbach-resonance could allow for spin selective interaction control and may induce instabilities in exotic pairing channels.

Feshbach resonances can also form in different scattering channels than the one considered here, i.e. inter-layer excitons in resonance with an intra-layer bound state, which could prove to be useful in the context of long-lived indirect exciton condensates~\cite{Wang19}. Furthermore, our work generates the opportunity to study few-body physics in two dimensional semiconductors. The tunable scattering length can be used to explore exotic multi-particle bound states, where a single electron binds multiple excitons~\cite{efimov_res_70,efimov_original_70}. While we specifically considered resonant scattering between excitons and electrons, Feshbach physics in 2D materials could be a generic phenomenon that may also be relevant for understanding purely electronic processes~\cite{slagle20}.

\begin{acknowledgments}
\textit{\textbf{Acknowledgments.---}} We thank I.~Schwartz and Y.~Shimazaki for discussions which motivated this work. The authors also thank A.~Cavalleri, F.~Helmrich and P.A.~Murthy for fruitful discussions. This work was supported by ETH Zurich. 
We also acknowledge support from the Technical University of Munich - Institute for Advanced Study, funded by the German Excellence Initiative and the European Union FP7 under grant agreement 291763, the Deutsche Forschungsgemeinschaft (DFG, German Research Foundation) under Germany’s Excellence Strategy--EXC--2111--390814868, TRR80 and DFG grant No. KN1254/1-2, from the European Research Council (ERC) under the European Union’s Horizon 2020 research and innovation programme (grant agreement No. 851161), as well as the priority program “Giant interactions in Rydberg systems”, DFG SPP 1929 GiRyd, Grant No. 428462134.
\end{acknowledgments}

%

\pagebreak
\clearpage

\onecolumngrid
\begin{center}
\textbf{\large Supplemental Material: Theory of Feshbach resonances in bilayer semiconductors}
\end{center}

\setcounter{equation}{0}
\setcounter{figure}{0}
\setcounter{table}{0}
\setcounter{page}{1}
\makeatletter
\renewcommand{\theequation}{S\arabic{equation}}
\renewcommand{\thefigure}{S\arabic{figure}}
\renewcommand{\bibnumfmt}[1]{[S#1]}
\renewcommand{\citenumfont}[1]{S#1}

\section{Renormalization of contact interactions}
While it is convenient to work with local two-body interactions of the form
\begin{equation}
\hat{V} = U \sum_{\mathbf{k},\mathbf{k'},\mathbf{q}} x^\dagger_\mathbf{k}x_{\mathbf{k}-\mathbf{q}} \, c^\dagger_\mathbf{k'} c_{\mathbf{k'}+\mathbf{q}},
\end{equation}
this leads to logarithmic UV divergences arising from high momentum virtual states. This is expected from contact-like models, which are only valid for low momentum scattering. To regularize the model we restrict loop momenta to lie below a cut-off scale $\Lambda$ and introduce a regularized interaction strength $U_\Lambda$. As with any effective theory, $U_\Lambda$ must be chosen to properly describe the physics of the underlying microscopic model. This is achieved by ensuring that the scattering length $a$, or the relevant binding energy $E_B^0$, is correctly reproduced. The binding energy of the exciton-electron complex in monolayer TMDs is known and was found to be $E_B^0 \simeq 30 \mathrm{meV}$. To fix $U_\Lambda$ we compute the T-matrix and match the position of its pole to the physical value of $E_B^0$
\begin{equation}
    \frac{1}{U_\Lambda} = \int\limits_{|\mathbf{k}|<\Lambda} \frac{d^2\mathbf{k}}{(2\pi)^2} \frac{1}{E^0_B - \frac{\mathbf{k}^2}{2\mu} + i 0^+}.
\end{equation}

\section{Non-equilibrium formalism}
In the following we find an approximate solution to Eq.~\ref{Eq:lindblad} of the main text using non-equilibirum field theory. It has been shown that generic Master equations can be mapped to a functional integral. This is achieved by rewriting the real time evolution of Eq.~\ref{Eq:lindblad} in a coherent state basis, which yields a field theory on a Keldysh contour~\cite{sieberer_16}. As the density matrix is now in general non-thermal, the system admits two possibly inequivalent correlation functions
\begin{equation}
    \begin{aligned}
    \mathcal{G}^>_\mathbf{k}(t,t') &= -i \langle x_\mathbf{k}(t) x^\dagger_\mathbf{k}(t')\rangle, \\ 
    \mathcal{G}^<_\mathbf{k}(t,t') &= -i \langle x^\dagger_\mathbf{k}(t') x_\mathbf{k}(t)\rangle,
    \end{aligned}
\end{equation}
which would be related to each other by fluctuation-dissipation relations if the system was in thermal equilibrium. In our case only the excitons couple to the radiation field, which leads to quantum jump operators $L_\mathbf{k}$, linear in the exciton operators ${x}_\mathbf{k}$ and ${x}^\dagger_\mathbf{k}$. Consequently, dissipation can be treated to all orders simply by modifying the impurity propagator. In the impurity limit $n_I= \sum_\mathbf{r}\langle x_\mathbf{r}^\dagger x_\mathbf{r}\rangle/V \approx 0$, this yields
\begin{equation}
\begin{aligned}
    \mathcal{G}^>(\omega,\mathbf{k})= \frac{2i \Gamma(\mathbf{k})}{(\omega-\mathbf{k}^2/2M)^2 + \Gamma(\mathbf{k})^2},\qquad \mathcal{G}^<(\omega,\mathbf{k})= 0.
\end{aligned}
    \label{Eq:diss_propagators}
\end{equation}
Other quantities, such as the retarded response function $\mathcal{G}^R$, contain no additional information and are defined in terms of $\mathcal{G}^\lessgtr$ as usual:
\begin{align*}
     \mathcal{G}^R(t,t') &= \Theta(t-t') \left( \mathcal{G}^> - \mathcal{G^<} \right)(t,t'),\\
      \mathcal{G}^{R/A}(\omega,\mathbf{k}) &= \frac{1}{\omega - \mathbf{k}^2/2M \pm i\Gamma(\mathbf{k})}.
\end{align*}
The electrons on the other hand are in thermal equilibrium and their correlation functions take the simpler form:
\begin{align*}
    G^>_\alpha(\omega,\mathbf{k})&=-2\pi i \; \left(1-n_F(\omega)\right) \delta (\omega - \mathbf{k}^2/2m-\varepsilon_\alpha+\epsilon_F),\qquad &G^<_\alpha(\omega,\mathbf{k})&= 2\pi i \; n_F(\omega) \delta (\omega - \mathbf{k}^2/2m-\varepsilon_\alpha+\epsilon_F),  \\ 
     G_\alpha^R(t,t') &= \Theta(t-t') \left( G^>_\alpha - G^<_\alpha \right)(t,t'),  \qquad
      &G_\alpha^{R}(\omega,\mathbf{k}) &= \frac{1}{\omega - \mathbf{k}^2/2m - \varepsilon_\alpha + i0^+},
\end{align*}
where $n_F(\omega)$ is the Fermi-Dirac distribution function and $\alpha \in \lbrace O,C\rbrace$ labels the channel.
 
\section{T-matrix approximation to the exciton self energy}
Here we determine the T-matrix of Eq.~\ref{Eq:lindblad} on a Keldysh contour, taking into account the ground state of the electrons. This will allow us to construct an approximate expression for the self energy. Although our nonequilibrium setting allows for additional correlations, the local-in-time structure of the interaction significantly restricts the number of independent components of the T-matrix~\cite{Babadi2013PhD}. In the end the T-matrix has the same causality structure as the propagators and can be expressed in terms of $T^\lessgtr(t,t')$. The diagrammatic structure remains is the same as for the two-particle problem in Eq.~\ref{Eq:tmatrix} except that time arguments now live on a Keldysh contour
\begin{fmffile}{tmatrix_2}
	\begin{align*}
	\begin{gathered}
	\begin{fmfgraph*}(60,40)
	\fmfleft{i1,i2}
	\fmfright{o1,o2}
	\fmf{fermion,fore=(0.3,,0.4,,0.8)}{i1,v}
	\fmf{fermion}{i2,v}
	\fmf{fermion,fore=(0.3,,0.4,,0.8)}{v,o1}
	\fmf{fermion}{v,o2}
	\fmfblob{.25w}{v}
	\end{fmfgraph*}
	\end{gathered}=
	\begin{gathered}
	\begin{fmfgraph*}(55,40)
	\fmfleft{i1,i2}
	\fmfright{o1,o2}
	\fmf{fermion,fore=(0.3,,0.4,,0.8)}{i1,v}
	\fmf{fermion}{i2,v}
	\fmf{fermion,fore=(0.3,,0.4,,0.8)}{v,o1}
	\fmf{fermion}{v,o2}
	\fmfdot{v}
	\end{fmfgraph*}
	\end{gathered}+
	\begin{gathered}
	\begin{fmfgraph*}(100,40)
	\fmfleft{i1,i2}
	\fmfright{o1,o2}
	\fmf{fermion,fore=(0.3,,0.4,,0.8)}{i1,v1}
	\fmf{fermion}{i2,v1}
	\fmf{fermion,fore=(0.3,,0.4,,0.8)}{v2,o1}
	\fmf{fermion}{v2,o2}
	\fmf{fermion,left=.5,tension=.5}{v1,v2}
	\fmf{fermion,right=.5,tension=.5,fore=(0.3,,0.4,,0.8)}{v1,v2}
	\fmfdot{v1}
	\fmfblob{.125w}{v2}
	\end{fmfgraph*}
	\end{gathered}.
	\end{align*}
\end{fmffile}
We use Langreth rules to decompose the above equation, which yields the following equations for the components of the T-matrix:
\begin{equation}
\begin{aligned}
     T_\mathbf{k}^\lessgtr(E) &= U_\Lambda \left[ K_\mathbf{k}^\lessgtr(E) T_\mathbf{k}^A(E) + K_\mathbf{k}^R(E)T_\mathbf{k}^\lessgtr(E)\right] &=& U_\Lambda\frac{K_\mathbf{k}^\lessgtr(E) T_\mathbf{k}^A(E)}{1-K_\mathbf{k}^R(E)},\\
     T_\mathbf{k}^{R/A}(E) & = U_\Lambda \left[\mathbb{1} + K^{R/A}_\mathbf{k}(E) T_\mathbf{k}^{R/A}(E)\right] &=& U_\Lambda \frac{1}{1- U_\Lambda K^{R/A}_\mathbf{k}(E)}, \qquad \mathrm{where}\\
     K^\lessgtr_{\alpha\beta}(t,\mathbf{x}) &= - \mathcal{G}^\lessgtr(t,\mathbf{x}) G^\lessgtr_{0,\alpha}(t,\mathbf{x})\delta_{\alpha\beta}, \quad \mathrm{and} \quad K^{R/A}_{\alpha\beta}(t,\mathbf{x})&=& \pm \Theta(\pm t)  \left(K^> - K^< \right)_{\alpha\beta}(t,\mathbf{x}). 
    \label{Eq:tmatrix_keldysh}
\end{aligned}
\end{equation}
$\Theta(x)$ is the Heaviside step function and the kernels $K$ can be explicitly computed as before. Eq.~\ref{Eq:tmatrix_keldysh} implies that $T^<(t,t') =0$, as it is proportional to $G^<$. This is a consequence of the impurity limit, as $T^<$ is proportional to the density of molecules, which vanishes for a single exciton.

We now compute an approximate self energy of the exciton from the many body T-matrix. Its form is given by the following Dyson equation
\begin{fmffile}{selfeng}
	\begin{align*}
	\begin{gathered}
	\begin{fmfgraph*}(40,20)
	\fmfleft{i}
	\fmfright{o}
	\fmf{fermion,fore=(0.15,,0.2,,0.4),width=1.8}{i,o}
	\end{fmfgraph*}
	\end{gathered}=
	\begin{gathered}
	\begin{fmfgraph*}(40,20)
	\fmfleft{i}
	\fmfright{o}
	\fmf{fermion,fore=(0.3,,0.4,,0.8)}{i,o}
	\end{fmfgraph*}
	\end{gathered}+
	\begin{gathered}
	\begin{fmfgraph}(80,15)
	\fmfleft{i}
	\fmfright{o}
	\fmf{phantom}{i,v1,v2,v3,v4,o}
	\fmf{fermion,fore=(0.3,,0.4,,0.8),tension=3.5}{i,v2}
	\fmf{fermion,fore=(0.15,,0.2,,0.4),tension=3.5,width=1.8}{v4,o}
	\fmf{fermion,right=4,tension=10}{v4,v2}
	\fmf{fermion}{v2,v3,v4}
	\fmfblob{.2w}{v3}
	\end{fmfgraph}	
	\end{gathered},
	\end{align*}
\end{fmffile}
which encodes the following equations
\begin{equation}
    \begin{aligned}
    \mathcal{G}^R(\omega,\mathbf{k}) &= \mathcal{G}^R_0(\omega,\mathbf{k}) + \mathcal{G}^R_0(\omega,\mathbf{k}) \cdot \Sigma^R(\omega,\mathbf{k})\cdot\mathcal{G}^R(\omega,
\mathbf{k}), \\
\Sigma^R(\omega,\mathbf{k})&= -i \int\frac{d^2q}{(2\pi)^2}\frac{d\delta\omega}{2\pi} \,\mathrm{Tr}\lbrace T^R_{\mathbf{k}+\mathbf{q}}(\omega+\delta\omega) \cdot G^<_0(\delta\omega,\mathbf{q})\rbrace,
    \end{aligned}
    \label{Eq:SDE}
\end{equation}
where the trace is performed over the two scattering channels. To obtain the second equations we have already assumed $T^<(\omega,\mathbf{k}) =0$, as the number of molecules is negligible in the impurity limit.

\begin{thebibliography}{50}%
\makeatletter
\providecommand \@ifxundefined [1]{%
 \@ifx{#1\undefined}
}%
\providecommand \@ifnum [1]{%
 \ifnum #1\expandafter \@firstoftwo
 \else \expandafter \@secondoftwo
 \fi
}%
\providecommand \@ifx [1]{%
 \ifx #1\expandafter \@firstoftwo
 \else \expandafter \@secondoftwo
 \fi
}%
\providecommand \natexlab [1]{#1}%
\providecommand \enquote  [1]{``#1''}%
\providecommand \bibnamefont  [1]{#1}%
\providecommand \bibfnamefont [1]{#1}%
\providecommand \citenamefont [1]{#1}%
\providecommand \href@noop [0]{\@secondoftwo}%
\providecommand \href [0]{\begingroup \@sanitize@url \@href}%
\providecommand \@href[1]{\@@startlink{#1}\@@href}%
\providecommand \@@href[1]{\endgroup#1\@@endlink}%
\providecommand \@sanitize@url [0]{\catcode `\\12\catcode `\$12\catcode
  `\&12\catcode `\#12\catcode `\^12\catcode `\_12\catcode `\%12\relax}%
\providecommand \@@startlink[1]{}%
\providecommand \@@endlink[0]{}%
\providecommand \url  [0]{\begingroup\@sanitize@url \@url }%
\providecommand \@url [1]{\endgroup\@href {#1}{\urlprefix }}%
\providecommand \urlprefix  [0]{URL }%
\providecommand \Eprint [0]{\href }%
\providecommand \doibase [0]{https://doi.org/}%
\providecommand \selectlanguage [0]{\@gobble}%
\providecommand \bibinfo  [0]{\@secondoftwo}%
\providecommand \bibfield  [0]{\@secondoftwo}%
\providecommand \translation [1]{[#1]}%
\providecommand \BibitemOpen [0]{}%
\providecommand \bibitemStop [0]{}%
\providecommand \bibitemNoStop [0]{.\EOS\space}%
\providecommand \EOS [0]{\spacefactor3000\relax}%
\providecommand \BibitemShut  [1]{\csname bibitem#1\endcsname}%
\let\auto@bib@innerbib\@empty
\bibitem [{\citenamefont {Andrei}\ \emph {et~al.}(2021)\citenamefont {Andrei},
  \citenamefont {Efetov}, \citenamefont {Jarillo-Herrero}, \citenamefont
  {MacDonald}, \citenamefont {Mak}, \citenamefont {Senthil}, \citenamefont
  {Tutuc}, \citenamefont {Yazdani},\ and\ \citenamefont {Young}}]{moire_rev}%
  \BibitemOpen
  \bibfield  {author} {\bibinfo {author} {\bibfnamefont {E.~Y.}\ \bibnamefont
  {Andrei}}, \bibinfo {author} {\bibfnamefont {D.~K.}\ \bibnamefont {Efetov}},
  \bibinfo {author} {\bibfnamefont {P.}~\bibnamefont {Jarillo-Herrero}},
  \bibinfo {author} {\bibfnamefont {A.~H.}\ \bibnamefont {MacDonald}}, \bibinfo
  {author} {\bibfnamefont {K.~F.}\ \bibnamefont {Mak}}, \bibinfo {author}
  {\bibfnamefont {T.}~\bibnamefont {Senthil}}, \bibinfo {author} {\bibfnamefont
  {E.}~\bibnamefont {Tutuc}}, \bibinfo {author} {\bibfnamefont
  {A.}~\bibnamefont {Yazdani}},\ and\ \bibinfo {author} {\bibfnamefont {A.~F.}\
  \bibnamefont {Young}},\ }\bibfield  {title} {\bibinfo {title} {The marvels of
  moir{\'e}materials},\ }\href {https://doi.org/10.1038/s41578-021-00284-1}
  {\bibfield  {journal} {\bibinfo  {journal} {Nature Reviews Materials}\
  }\textbf {\bibinfo {volume} {6}},\ \bibinfo {pages} {201} (\bibinfo {year}
  {2021})}\BibitemShut {NoStop}%
\bibitem [{\citenamefont {Liu}\ \emph {et~al.}(2020)\citenamefont {Liu},
  \citenamefont {Li}, \citenamefont {Watanabe}, \citenamefont {Taniguchi},
  \citenamefont {Hone}, \citenamefont {Halperin}, \citenamefont {Kim},\ and\
  \citenamefont {Dean}}]{liu20_bilayer}%
  \BibitemOpen
  \bibfield  {author} {\bibinfo {author} {\bibfnamefont {X.}~\bibnamefont
  {Liu}}, \bibinfo {author} {\bibfnamefont {J.~I.~A.}\ \bibnamefont {Li}},
  \bibinfo {author} {\bibfnamefont {K.}~\bibnamefont {Watanabe}}, \bibinfo
  {author} {\bibfnamefont {T.}~\bibnamefont {Taniguchi}}, \bibinfo {author}
  {\bibfnamefont {J.}~\bibnamefont {Hone}}, \bibinfo {author} {\bibfnamefont
  {B.~I.}\ \bibnamefont {Halperin}}, \bibinfo {author} {\bibfnamefont
  {P.}~\bibnamefont {Kim}},\ and\ \bibinfo {author} {\bibfnamefont {C.~R.}\
  \bibnamefont {Dean}},\ }\href@noop {} {\bibinfo {title} {Crossover between
  strongly-coupled and weakly-coupled exciton superfluids}} (\bibinfo {year}
  {2020}),\ \Eprint {https://arxiv.org/abs/2012.05916} {arXiv:2012.05916
  [cond-mat.mes-hall]} \BibitemShut {NoStop}%
\bibitem [{\citenamefont {Cao}\ \emph {et~al.}(2018)\citenamefont {Cao},
  \citenamefont {Fatemi}, \citenamefont {Fang}, \citenamefont {Watanabe},
  \citenamefont {Taniguchi}, \citenamefont {Kaxiras},\ and\ \citenamefont
  {Jarillo-Herrero}}]{cao_18}%
  \BibitemOpen
  \bibfield  {author} {\bibinfo {author} {\bibfnamefont {Y.}~\bibnamefont
  {Cao}}, \bibinfo {author} {\bibfnamefont {V.}~\bibnamefont {Fatemi}},
  \bibinfo {author} {\bibfnamefont {S.}~\bibnamefont {Fang}}, \bibinfo {author}
  {\bibfnamefont {K.}~\bibnamefont {Watanabe}}, \bibinfo {author}
  {\bibfnamefont {T.}~\bibnamefont {Taniguchi}}, \bibinfo {author}
  {\bibfnamefont {E.}~\bibnamefont {Kaxiras}},\ and\ \bibinfo {author}
  {\bibfnamefont {P.}~\bibnamefont {Jarillo-Herrero}},\ }\bibfield  {title}
  {\bibinfo {title} {Unconventional superconductivity in magic-angle graphene
  superlattices},\ }\href {https://doi.org/10.1038/nature26160} {\bibfield
  {journal} {\bibinfo  {journal} {Nature}\ }\textbf {\bibinfo {volume} {556}},\
  \bibinfo {pages} {43} (\bibinfo {year} {2018})}\BibitemShut {NoStop}%
\bibitem [{\citenamefont {Bistritzer}\ and\ \citenamefont
  {MacDonald}(2011)}]{bistritzer11}%
  \BibitemOpen
  \bibfield  {author} {\bibinfo {author} {\bibfnamefont {R.}~\bibnamefont
  {Bistritzer}}\ and\ \bibinfo {author} {\bibfnamefont {A.~H.}\ \bibnamefont
  {MacDonald}},\ }\bibfield  {title} {\bibinfo {title} {Moir{\'e} bands in
  twisted double-layer graphene},\ }\href
  {https://doi.org/10.1073/pnas.1108174108} {\bibfield  {journal} {\bibinfo
  {journal} {Proceedings of the National Academy of Sciences}\ }\textbf
  {\bibinfo {volume} {108}},\ \bibinfo {pages} {12233} (\bibinfo {year}
  {2011})}\BibitemShut {NoStop}%
\bibitem [{\citenamefont {Regan}\ \emph {et~al.}(2020)\citenamefont {Regan},
  \citenamefont {Wang}, \citenamefont {Jin}, \citenamefont {Bakti~Utama},
  \citenamefont {Gao}, \citenamefont {Wei}, \citenamefont {Zhao}, \citenamefont
  {Zhao}, \citenamefont {Zhang}, \citenamefont {Yumigeta}, \citenamefont
  {Blei}, \citenamefont {Carlstr{\"o}m}, \citenamefont {Watanabe},
  \citenamefont {Taniguchi}, \citenamefont {Tongay}, \citenamefont {Crommie},
  \citenamefont {Zettl},\ and\ \citenamefont {Wang}}]{regan_20}%
  \BibitemOpen
  \bibfield  {author} {\bibinfo {author} {\bibfnamefont {E.~C.}\ \bibnamefont
  {Regan}}, \bibinfo {author} {\bibfnamefont {D.}~\bibnamefont {Wang}},
  \bibinfo {author} {\bibfnamefont {C.}~\bibnamefont {Jin}}, \bibinfo {author}
  {\bibfnamefont {M.~I.}\ \bibnamefont {Bakti~Utama}}, \bibinfo {author}
  {\bibfnamefont {B.}~\bibnamefont {Gao}}, \bibinfo {author} {\bibfnamefont
  {X.}~\bibnamefont {Wei}}, \bibinfo {author} {\bibfnamefont {S.}~\bibnamefont
  {Zhao}}, \bibinfo {author} {\bibfnamefont {W.}~\bibnamefont {Zhao}}, \bibinfo
  {author} {\bibfnamefont {Z.}~\bibnamefont {Zhang}}, \bibinfo {author}
  {\bibfnamefont {K.}~\bibnamefont {Yumigeta}}, \bibinfo {author}
  {\bibfnamefont {M.}~\bibnamefont {Blei}}, \bibinfo {author} {\bibfnamefont
  {J.~D.}\ \bibnamefont {Carlstr{\"o}m}}, \bibinfo {author} {\bibfnamefont
  {K.}~\bibnamefont {Watanabe}}, \bibinfo {author} {\bibfnamefont
  {T.}~\bibnamefont {Taniguchi}}, \bibinfo {author} {\bibfnamefont
  {S.}~\bibnamefont {Tongay}}, \bibinfo {author} {\bibfnamefont
  {M.}~\bibnamefont {Crommie}}, \bibinfo {author} {\bibfnamefont
  {A.}~\bibnamefont {Zettl}},\ and\ \bibinfo {author} {\bibfnamefont
  {F.}~\bibnamefont {Wang}},\ }\bibfield  {title} {\bibinfo {title} {{Mott and
  generalized Wigner crystal states in WSe2/WS2 moir{\'e} superlattices}},\
  }\href {https://doi.org/10.1038/s41586-020-2092-4} {\bibfield  {journal}
  {\bibinfo  {journal} {Nature}\ }\textbf {\bibinfo {volume} {579}},\ \bibinfo
  {pages} {359} (\bibinfo {year} {2020})}\BibitemShut {NoStop}%
\bibitem [{\citenamefont {Tang}\ \emph {et~al.}(2020)\citenamefont {Tang},
  \citenamefont {Li}, \citenamefont {Li}, \citenamefont {Xu}, \citenamefont
  {Liu}, \citenamefont {Barmak}, \citenamefont {Watanabe}, \citenamefont
  {Taniguchi}, \citenamefont {MacDonald}, \citenamefont {Shan},\ and\
  \citenamefont {Mak}}]{tang_20}%
  \BibitemOpen
  \bibfield  {author} {\bibinfo {author} {\bibfnamefont {Y.}~\bibnamefont
  {Tang}}, \bibinfo {author} {\bibfnamefont {L.}~\bibnamefont {Li}}, \bibinfo
  {author} {\bibfnamefont {T.}~\bibnamefont {Li}}, \bibinfo {author}
  {\bibfnamefont {Y.}~\bibnamefont {Xu}}, \bibinfo {author} {\bibfnamefont
  {S.}~\bibnamefont {Liu}}, \bibinfo {author} {\bibfnamefont {K.}~\bibnamefont
  {Barmak}}, \bibinfo {author} {\bibfnamefont {K.}~\bibnamefont {Watanabe}},
  \bibinfo {author} {\bibfnamefont {T.}~\bibnamefont {Taniguchi}}, \bibinfo
  {author} {\bibfnamefont {A.~H.}\ \bibnamefont {MacDonald}}, \bibinfo {author}
  {\bibfnamefont {J.}~\bibnamefont {Shan}},\ and\ \bibinfo {author}
  {\bibfnamefont {K.~F.}\ \bibnamefont {Mak}},\ }\bibfield  {title} {\bibinfo
  {title} {{Simulation of Hubbard model physics in WSe2/WS2 moir{\'e}
  superlattices}},\ }\href {https://doi.org/10.1038/s41586-020-2085-3}
  {\bibfield  {journal} {\bibinfo  {journal} {Nature}\ }\textbf {\bibinfo
  {volume} {579}},\ \bibinfo {pages} {353} (\bibinfo {year}
  {2020})}\BibitemShut {NoStop}%
\bibitem [{\citenamefont {Shimazaki}\ \emph {et~al.}(2020)\citenamefont
  {Shimazaki}, \citenamefont {Schwartz}, \citenamefont {Watanabe},
  \citenamefont {Taniguchi}, \citenamefont {Kroner},\ and\ \citenamefont
  {Imamo{\u g}lu}}]{yuya_20}%
  \BibitemOpen
  \bibfield  {author} {\bibinfo {author} {\bibfnamefont {Y.}~\bibnamefont
  {Shimazaki}}, \bibinfo {author} {\bibfnamefont {I.}~\bibnamefont {Schwartz}},
  \bibinfo {author} {\bibfnamefont {K.}~\bibnamefont {Watanabe}}, \bibinfo
  {author} {\bibfnamefont {T.}~\bibnamefont {Taniguchi}}, \bibinfo {author}
  {\bibfnamefont {M.}~\bibnamefont {Kroner}},\ and\ \bibinfo {author}
  {\bibfnamefont {A.}~\bibnamefont {Imamo{\u g}lu}},\ }\bibfield  {title}
  {\bibinfo {title} {{Strongly correlated electrons and hybrid excitons in a
  moir{\'e} heterostructure}},\ }\href
  {https://doi.org/10.1038/s41586-020-2191-2} {\bibfield  {journal} {\bibinfo
  {journal} {Nature}\ }\textbf {\bibinfo {volume} {580}},\ \bibinfo {pages}
  {472} (\bibinfo {year} {2020})}\BibitemShut {NoStop}%
\bibitem [{\citenamefont {Ferrier-Barbut}\ \emph {et~al.}(2014)\citenamefont
  {Ferrier-Barbut}, \citenamefont {Delehaye}, \citenamefont {Laurent},
  \citenamefont {Grier}, \citenamefont {Pierce}, \citenamefont {Rem},
  \citenamefont {Chevy},\ and\ \citenamefont {Salomon}}]{barbut15}%
  \BibitemOpen
  \bibfield  {author} {\bibinfo {author} {\bibfnamefont {I.}~\bibnamefont
  {Ferrier-Barbut}}, \bibinfo {author} {\bibfnamefont {M.}~\bibnamefont
  {Delehaye}}, \bibinfo {author} {\bibfnamefont {S.}~\bibnamefont {Laurent}},
  \bibinfo {author} {\bibfnamefont {A.~T.}\ \bibnamefont {Grier}}, \bibinfo
  {author} {\bibfnamefont {M.}~\bibnamefont {Pierce}}, \bibinfo {author}
  {\bibfnamefont {B.~S.}\ \bibnamefont {Rem}}, \bibinfo {author} {\bibfnamefont
  {F.}~\bibnamefont {Chevy}},\ and\ \bibinfo {author} {\bibfnamefont
  {C.}~\bibnamefont {Salomon}},\ }\bibfield  {title} {\bibinfo {title} {A
  mixture of bose and fermi superfluids},\ }\href
  {https://doi.org/10.1126/science.1255380} {\bibfield  {journal} {\bibinfo
  {journal} {Science}\ }\textbf {\bibinfo {volume} {345}},\ \bibinfo {pages}
  {1035} (\bibinfo {year} {2014})}\BibitemShut {NoStop}%
\bibitem [{\citenamefont {DeSalvo}\ \emph {et~al.}(2019)\citenamefont
  {DeSalvo}, \citenamefont {Patel}, \citenamefont {Cai},\ and\ \citenamefont
  {Chin}}]{desalvo19}%
  \BibitemOpen
  \bibfield  {author} {\bibinfo {author} {\bibfnamefont {B.~J.}\ \bibnamefont
  {DeSalvo}}, \bibinfo {author} {\bibfnamefont {K.}~\bibnamefont {Patel}},
  \bibinfo {author} {\bibfnamefont {G.}~\bibnamefont {Cai}},\ and\ \bibinfo
  {author} {\bibfnamefont {C.}~\bibnamefont {Chin}},\ }\bibfield  {title}
  {\bibinfo {title} {Observation of fermion-mediated interactions between
  bosonic atoms},\ }\href {https://doi.org/10.1038/s41586-019-1055-0}
  {\bibfield  {journal} {\bibinfo  {journal} {Nature}\ }\textbf {\bibinfo
  {volume} {568}},\ \bibinfo {pages} {61} (\bibinfo {year} {2019})}\BibitemShut
  {NoStop}%
\bibitem [{\citenamefont {Fritsche}\ \emph {et~al.}(2021)\citenamefont
  {Fritsche}, \citenamefont {Baroni}, \citenamefont {Dobler}, \citenamefont
  {Kirilov}, \citenamefont {Huang}, \citenamefont {Grimm}, \citenamefont
  {Bruun},\ and\ \citenamefont {Massignan}}]{fritsche21}%
  \BibitemOpen
  \bibfield  {author} {\bibinfo {author} {\bibfnamefont {I.}~\bibnamefont
  {Fritsche}}, \bibinfo {author} {\bibfnamefont {C.}~\bibnamefont {Baroni}},
  \bibinfo {author} {\bibfnamefont {E.}~\bibnamefont {Dobler}}, \bibinfo
  {author} {\bibfnamefont {E.}~\bibnamefont {Kirilov}}, \bibinfo {author}
  {\bibfnamefont {B.}~\bibnamefont {Huang}}, \bibinfo {author} {\bibfnamefont
  {R.}~\bibnamefont {Grimm}}, \bibinfo {author} {\bibfnamefont {G.~M.}\
  \bibnamefont {Bruun}},\ and\ \bibinfo {author} {\bibfnamefont
  {P.}~\bibnamefont {Massignan}},\ }\bibfield  {title} {\bibinfo {title}
  {Stability and breakdown of fermi polarons in a strongly interacting
  fermi-bose mixture},\ }\href {https://arxiv.org/abs/2103.03625} {\bibfield
  {journal} {\bibinfo  {journal} {arXiv preprint arXiv:2103.03625}\ } (\bibinfo
  {year} {2021})}\BibitemShut {NoStop}%
\bibitem [{\citenamefont {Feshbach}(1962)}]{feshbach1962}%
  \BibitemOpen
  \bibfield  {author} {\bibinfo {author} {\bibfnamefont {H.}~\bibnamefont
  {Feshbach}},\ }\bibfield  {title} {\bibinfo {title} {A unified theory of
  nuclear reactions. ii},\ }\href
  {https://doi.org/https://doi.org/10.1016/0003-4916(62)90221-X} {\bibfield
  {journal} {\bibinfo  {journal} {Annals of Physics}\ }\textbf {\bibinfo
  {volume} {19}},\ \bibinfo {pages} {287} (\bibinfo {year} {1962})}\BibitemShut
  {NoStop}%
\bibitem [{\citenamefont {Chin}\ \emph {et~al.}(2010)\citenamefont {Chin},
  \citenamefont {Grimm}, \citenamefont {Julienne},\ and\ \citenamefont
  {Tiesinga}}]{chin_rev_10}%
  \BibitemOpen
  \bibfield  {author} {\bibinfo {author} {\bibfnamefont {C.}~\bibnamefont
  {Chin}}, \bibinfo {author} {\bibfnamefont {R.}~\bibnamefont {Grimm}},
  \bibinfo {author} {\bibfnamefont {P.}~\bibnamefont {Julienne}},\ and\
  \bibinfo {author} {\bibfnamefont {E.}~\bibnamefont {Tiesinga}},\ }\bibfield
  {title} {\bibinfo {title} {Feshbach resonances in ultracold gases},\ }\href
  {https://doi.org/10.1103/RevModPhys.82.1225} {\bibfield  {journal} {\bibinfo
  {journal} {Rev. Mod. Phys.}\ }\textbf {\bibinfo {volume} {82}},\ \bibinfo
  {pages} {1225} (\bibinfo {year} {2010})}\BibitemShut {NoStop}%
\bibitem [{\citenamefont {Ketterle}\ and\ \citenamefont
  {Zwierlein}(2008)}]{ketterle_08}%
  \BibitemOpen
  \bibfield  {author} {\bibinfo {author} {\bibfnamefont {W.}~\bibnamefont
  {Ketterle}}\ and\ \bibinfo {author} {\bibfnamefont {M.~W.}\ \bibnamefont
  {Zwierlein}},\ }\bibfield  {title} {\bibinfo {title} {Making, probing and
  understanding ultracold fermi gases},\ }\href
  {https://doi.org/10.1393/ncr/i2008-10033-1} {\bibfield  {journal} {\bibinfo
  {journal} {Nuovo Cimento Rivista Serie}\ }\textbf {\bibinfo {volume} {31}},\
  \bibinfo {pages} {247} (\bibinfo {year} {2008})}\BibitemShut {NoStop}%
\bibitem [{\citenamefont {Bloch}\ \emph {et~al.}(2008)\citenamefont {Bloch},
  \citenamefont {Dalibard},\ and\ \citenamefont {Zwerger}}]{zwerger_rev_08}%
  \BibitemOpen
  \bibfield  {author} {\bibinfo {author} {\bibfnamefont {I.}~\bibnamefont
  {Bloch}}, \bibinfo {author} {\bibfnamefont {J.}~\bibnamefont {Dalibard}},\
  and\ \bibinfo {author} {\bibfnamefont {W.}~\bibnamefont {Zwerger}},\
  }\bibfield  {title} {\bibinfo {title} {Many-body physics with ultracold
  gases},\ }\href {https://doi.org/10.1103/RevModPhys.80.885} {\bibfield
  {journal} {\bibinfo  {journal} {Rev. Mod. Phys.}\ }\textbf {\bibinfo {volume}
  {80}},\ \bibinfo {pages} {885} (\bibinfo {year} {2008})}\BibitemShut
  {NoStop}%
\bibitem [{\citenamefont {Duine}\ and\ \citenamefont {Stoof}(2004)}]{duine_04}%
  \BibitemOpen
  \bibfield  {author} {\bibinfo {author} {\bibfnamefont {R.}~\bibnamefont
  {Duine}}\ and\ \bibinfo {author} {\bibfnamefont {H.}~\bibnamefont {Stoof}},\
  }\bibfield  {title} {\bibinfo {title} {Atom–molecule coherence in bose
  gases},\ }\href
  {https://doi.org/https://doi.org/10.1016/j.physrep.2004.03.003} {\bibfield
  {journal} {\bibinfo  {journal} {Physics Reports}\ }\textbf {\bibinfo {volume}
  {396}},\ \bibinfo {pages} {115} (\bibinfo {year} {2004})}\BibitemShut
  {NoStop}%
\bibitem [{\citenamefont {Schwartz}\ \emph {et~al.}(2021)\citenamefont
  {Schwartz}, \citenamefont {Shimazaki}, \citenamefont {Kuhlenkamp},
  \citenamefont {Watanabe}, \citenamefont {Taniguchi}, \citenamefont {Kroner},\
  and\ \citenamefont {Imamo{\u g}lu}}]{schwartz21}%
  \BibitemOpen
  \bibfield  {author} {\bibinfo {author} {\bibfnamefont {I.}~\bibnamefont
  {Schwartz}}, \bibinfo {author} {\bibfnamefont {Y.}~\bibnamefont {Shimazaki}},
  \bibinfo {author} {\bibfnamefont {C.}~\bibnamefont {Kuhlenkamp}}, \bibinfo
  {author} {\bibfnamefont {K.}~\bibnamefont {Watanabe}}, \bibinfo {author}
  {\bibfnamefont {T.}~\bibnamefont {Taniguchi}}, \bibinfo {author}
  {\bibfnamefont {M.}~\bibnamefont {Kroner}},\ and\ \bibinfo {author}
  {\bibfnamefont {A.}~\bibnamefont {Imamo{\u g}lu}},\ }\bibfield  {title}
  {\bibinfo {title} {{In preparation}},\ }\href@noop {} {\  (\bibinfo {year}
  {2021})}\BibitemShut {NoStop}%
\bibitem [{Note4()}]{Note4}%
  \BibitemOpen
  \bibinfo {note} {Large electric fields could overcome the energy difference
  resulting in sizable hybridization with inter-layer excitons, which would
  then contribute to inter-layer scattering of electrons and excitons. Here are
  we are interested in smaller fields on the order of the trion binding energy
  where these processes are suppressed.}\BibitemShut {Stop}%
\bibitem [{\citenamefont {Fey}\ \emph {et~al.}(2020)\citenamefont {Fey},
  \citenamefont {Schmelcher}, \citenamefont {Imamoglu},\ and\ \citenamefont
  {Schmidt}}]{Fey20}%
  \BibitemOpen
  \bibfield  {author} {\bibinfo {author} {\bibfnamefont {C.}~\bibnamefont
  {Fey}}, \bibinfo {author} {\bibfnamefont {P.}~\bibnamefont {Schmelcher}},
  \bibinfo {author} {\bibfnamefont {A.}~\bibnamefont {Imamoglu}},\ and\
  \bibinfo {author} {\bibfnamefont {R.}~\bibnamefont {Schmidt}},\ }\bibfield
  {title} {\bibinfo {title} {Theory of exciton-electron scattering in
  atomically thin semiconductors},\ }\href
  {https://doi.org/10.1103/PhysRevB.101.195417} {\bibfield  {journal} {\bibinfo
   {journal} {Phys. Rev. B}\ }\textbf {\bibinfo {volume} {101}},\ \bibinfo
  {pages} {195417} (\bibinfo {year} {2020})}\BibitemShut {NoStop}%
\bibitem [{\citenamefont {Adhikari}(1986)}]{adhikari}%
  \BibitemOpen
  \bibfield  {author} {\bibinfo {author} {\bibfnamefont {S.~K.}\ \bibnamefont
  {Adhikari}},\ }\bibfield  {title} {\bibinfo {title} {Quantum scattering in
  two dimensions},\ }\href {https://doi.org/10.1119/1.14623} {\bibfield
  {journal} {\bibinfo  {journal} {American Journal of Physics}\ }\textbf
  {\bibinfo {volume} {54}},\ \bibinfo {pages} {362} (\bibinfo {year} {1986})},\
  \Eprint {https://arxiv.org/abs/https://doi.org/10.1119/1.14623}
  {https://doi.org/10.1119/1.14623} \BibitemShut {NoStop}%
\bibitem [{\citenamefont {Weinberg}(1995)}]{Weinberg}%
  \BibitemOpen
  \bibfield  {author} {\bibinfo {author} {\bibfnamefont {S.}~\bibnamefont
  {Weinberg}},\ }\href {https://doi.org/10.1017/CBO9781139644167} {\emph
  {\bibinfo {title} {The Quantum Theory of Fields}}},\ Vol.~\bibinfo {volume}
  {1}\ (\bibinfo  {publisher} {Cambridge University Press},\ \bibinfo {year}
  {1995})\BibitemShut {NoStop}%
\bibitem [{Note1()}]{Note1}%
  \BibitemOpen
  \bibinfo {note} {More importantly, for any given electron density, it is
  possible to choose $E_z$ such that $k_F a_O \sim 1$, which yields the
  strong correlation regime.}\BibitemShut {Stop}%
\bibitem [{\citenamefont {Kanjilal}\ and\ \citenamefont
  {Blume}(2006)}]{kanjilal06}%
  \BibitemOpen
  \bibfield  {author} {\bibinfo {author} {\bibfnamefont {K.}~\bibnamefont
  {Kanjilal}}\ and\ \bibinfo {author} {\bibfnamefont {D.}~\bibnamefont
  {Blume}},\ }\bibfield  {title} {\bibinfo {title} {Coupled-channel
  pseudopotential description of the feshbach resonance in two dimensions},\
  }\href {https://doi.org/10.1103/PhysRevA.73.060701} {\bibfield  {journal}
  {\bibinfo  {journal} {Phys. Rev. A}\ }\textbf {\bibinfo {volume} {73}},\
  \bibinfo {pages} {060701} (\bibinfo {year} {2006})}\BibitemShut {NoStop}%
\bibitem [{\citenamefont {Sidler}\ \emph {et~al.}(2017)\citenamefont {Sidler},
  \citenamefont {Back}, \citenamefont {Cotlet}, \citenamefont {Srivastava},
  \citenamefont {Fink}, \citenamefont {Kroner}, \citenamefont {Demler},\ and\
  \citenamefont {Imamoglu}}]{sidler_polaron}%
  \BibitemOpen
  \bibfield  {author} {\bibinfo {author} {\bibfnamefont {M.}~\bibnamefont
  {Sidler}}, \bibinfo {author} {\bibfnamefont {P.}~\bibnamefont {Back}},
  \bibinfo {author} {\bibfnamefont {O.}~\bibnamefont {Cotlet}}, \bibinfo
  {author} {\bibfnamefont {A.}~\bibnamefont {Srivastava}}, \bibinfo {author}
  {\bibfnamefont {T.}~\bibnamefont {Fink}}, \bibinfo {author} {\bibfnamefont
  {M.}~\bibnamefont {Kroner}}, \bibinfo {author} {\bibfnamefont
  {E.}~\bibnamefont {Demler}},\ and\ \bibinfo {author} {\bibfnamefont
  {A.}~\bibnamefont {Imamoglu}},\ }\bibfield  {title} {\bibinfo {title} {Fermi
  polaron-polaritons in charge-tunable atomically thin semiconductors},\ }\href
  {https://doi.org/10.1038/nphys3949} {\bibfield  {journal} {\bibinfo
  {journal} {Nature Physics}\ }\textbf {\bibinfo {volume} {13}},\ \bibinfo
  {pages} {255} (\bibinfo {year} {2017})}\BibitemShut {NoStop}%
\bibitem [{\citenamefont {Suris}(2003)}]{suris_03}%
  \BibitemOpen
  \bibfield  {author} {\bibinfo {author} {\bibfnamefont {R.~A.}\ \bibnamefont
  {Suris}},\ }\bibfield  {title} {\bibinfo {title} {Correlation between trion
  and hole in fermi distribution in process of trion photo-excitation in doped
  qws},\ }in\ \href {https://doi.org/10.1007/978-94-010-0078-9_9} {\emph
  {\bibinfo {booktitle} {Optical Properties of 2D Systems with Interacting
  Electrons}}},\ \bibinfo {editor} {edited by\ \bibinfo {editor} {\bibfnamefont
  {W.~J.}\ \bibnamefont {Ossau}}\ and\ \bibinfo {editor} {\bibfnamefont
  {R.}~\bibnamefont {Suris}}}\ (\bibinfo  {publisher} {Springer Netherlands},\
  \bibinfo {address} {Dordrecht},\ \bibinfo {year} {2003})\ pp.\ \bibinfo
  {pages} {111--124}\BibitemShut {NoStop}%
\bibitem [{\citenamefont {Efimkin}\ and\ \citenamefont
  {MacDonald}(2017)}]{efimkin}%
  \BibitemOpen
  \bibfield  {author} {\bibinfo {author} {\bibfnamefont {D.~K.}\ \bibnamefont
  {Efimkin}}\ and\ \bibinfo {author} {\bibfnamefont {A.~H.}\ \bibnamefont
  {MacDonald}},\ }\bibfield  {title} {\bibinfo {title} {Many-body theory of
  trion absorption features in two-dimensional semiconductors},\ }\href
  {https://doi.org/10.1103/PhysRevB.95.035417} {\bibfield  {journal} {\bibinfo
  {journal} {Phys. Rev. B}\ }\textbf {\bibinfo {volume} {95}},\ \bibinfo
  {pages} {035417} (\bibinfo {year} {2017})}\BibitemShut {NoStop}%
\bibitem [{Note2()}]{Note2}%
  \BibitemOpen
  \bibinfo {note} {Finite momentum excitons are long-lived due to the steep
  light-cone, which renders only $\protect \mathbf {k}=0$ excitons optically
  active.}\BibitemShut {Stop}%
\bibitem [{\citenamefont {Fumi}(1955)}]{fumi_thm}%
  \BibitemOpen
  \bibfield  {author} {\bibinfo {author} {\bibfnamefont {F.}~\bibnamefont
  {Fumi}},\ }\bibfield  {title} {\bibinfo {title} {Cxvi. vacancies in
  monovalent metals},\ }\href {https://doi.org/10.1080/14786440908520622}
  {\bibfield  {journal} {\bibinfo  {journal} {The London, Edinburgh, and Dublin
  Philosophical Magazine and Journal of Science}\ }\textbf {\bibinfo {volume}
  {46}},\ \bibinfo {pages} {1007} (\bibinfo {year} {1955})}\BibitemShut
  {NoStop}%
\bibitem [{\citenamefont {Schmidt}\ \emph {et~al.}(2018)\citenamefont
  {Schmidt}, \citenamefont {Knap}, \citenamefont {Ivanov}, \citenamefont {You},
  \citenamefont {Cetina},\ and\ \citenamefont {Demler}}]{schmidt_rev_2018}%
  \BibitemOpen
  \bibfield  {author} {\bibinfo {author} {\bibfnamefont {R.}~\bibnamefont
  {Schmidt}}, \bibinfo {author} {\bibfnamefont {M.}~\bibnamefont {Knap}},
  \bibinfo {author} {\bibfnamefont {D.~A.}\ \bibnamefont {Ivanov}}, \bibinfo
  {author} {\bibfnamefont {J.-S.}\ \bibnamefont {You}}, \bibinfo {author}
  {\bibfnamefont {M.}~\bibnamefont {Cetina}},\ and\ \bibinfo {author}
  {\bibfnamefont {E.}~\bibnamefont {Demler}},\ }\bibfield  {title} {\bibinfo
  {title} {Universal many-body response of heavy impurities coupled to a fermi
  sea: a review of recent progress},\ }\href
  {https://doi.org/10.1088/1361-6633/aa9593} {\bibfield  {journal} {\bibinfo
  {journal} {Rep. Prog. Phys.}\ }\textbf {\bibinfo {volume} {81}},\ \bibinfo
  {pages} {024401} (\bibinfo {year} {2018})}\BibitemShut {NoStop}%
\bibitem [{\citenamefont {Chevy}(2006)}]{chevy}%
  \BibitemOpen
  \bibfield  {author} {\bibinfo {author} {\bibfnamefont {F.}~\bibnamefont
  {Chevy}},\ }\bibfield  {title} {\bibinfo {title} {Universal phase diagram of
  a strongly interacting fermi gas with unbalanced spin populations},\ }\href
  {https://doi.org/10.1103/PhysRevA.74.063628} {\bibfield  {journal} {\bibinfo
  {journal} {Phys. Rev. A}\ }\textbf {\bibinfo {volume} {74}},\ \bibinfo
  {pages} {063628} (\bibinfo {year} {2006})}\BibitemShut {NoStop}%
\bibitem [{\citenamefont {Combescot}\ \emph {et~al.}(2007)\citenamefont
  {Combescot}, \citenamefont {Recati}, \citenamefont {Lobo},\ and\
  \citenamefont {Chevy}}]{combescot}%
  \BibitemOpen
  \bibfield  {author} {\bibinfo {author} {\bibfnamefont {R.}~\bibnamefont
  {Combescot}}, \bibinfo {author} {\bibfnamefont {A.}~\bibnamefont {Recati}},
  \bibinfo {author} {\bibfnamefont {C.}~\bibnamefont {Lobo}},\ and\ \bibinfo
  {author} {\bibfnamefont {F.}~\bibnamefont {Chevy}},\ }\bibfield  {title}
  {\bibinfo {title} {Normal state of highly polarized fermi gases: Simple
  many-body approaches},\ }\href
  {https://doi.org/10.1103/PhysRevLett.98.180402} {\bibfield  {journal}
  {\bibinfo  {journal} {Phys. Rev. Lett.}\ }\textbf {\bibinfo {volume} {98}},\
  \bibinfo {pages} {180402} (\bibinfo {year} {2007})}\BibitemShut {NoStop}%
\bibitem [{\citenamefont {Schmidt}\ \emph {et~al.}(2012)\citenamefont
  {Schmidt}, \citenamefont {Enss}, \citenamefont {Pietil\"a},\ and\
  \citenamefont {Demler}}]{schmidt_polaron2d}%
  \BibitemOpen
  \bibfield  {author} {\bibinfo {author} {\bibfnamefont {R.}~\bibnamefont
  {Schmidt}}, \bibinfo {author} {\bibfnamefont {T.}~\bibnamefont {Enss}},
  \bibinfo {author} {\bibfnamefont {V.}~\bibnamefont {Pietil\"a}},\ and\
  \bibinfo {author} {\bibfnamefont {E.}~\bibnamefont {Demler}},\ }\bibfield
  {title} {\bibinfo {title} {Fermi polarons in two dimensions},\ }\href
  {https://doi.org/10.1103/PhysRevA.85.021602} {\bibfield  {journal} {\bibinfo
  {journal} {Phys. Rev. A}\ }\textbf {\bibinfo {volume} {85}},\ \bibinfo
  {pages} {021602} (\bibinfo {year} {2012})}\BibitemShut {NoStop}%
\bibitem [{\citenamefont {Vlietinck}\ \emph {et~al.}(2014)\citenamefont
  {Vlietinck}, \citenamefont {Ryckebusch},\ and\ \citenamefont
  {Van~Houcke}}]{qmc_1}%
  \BibitemOpen
  \bibfield  {author} {\bibinfo {author} {\bibfnamefont {J.}~\bibnamefont
  {Vlietinck}}, \bibinfo {author} {\bibfnamefont {J.}~\bibnamefont
  {Ryckebusch}},\ and\ \bibinfo {author} {\bibfnamefont {K.}~\bibnamefont
  {Van~Houcke}},\ }\bibfield  {title} {\bibinfo {title} {Diagrammatic monte
  carlo study of the fermi polaron in two dimensions},\ }\href
  {https://doi.org/10.1103/PhysRevB.89.085119} {\bibfield  {journal} {\bibinfo
  {journal} {Phys. Rev. B}\ }\textbf {\bibinfo {volume} {89}},\ \bibinfo
  {pages} {085119} (\bibinfo {year} {2014})}\BibitemShut {NoStop}%
\bibitem [{\citenamefont {Kroiss}\ and\ \citenamefont {Pollet}(2014)}]{qmc_2}%
  \BibitemOpen
  \bibfield  {author} {\bibinfo {author} {\bibfnamefont {P.}~\bibnamefont
  {Kroiss}}\ and\ \bibinfo {author} {\bibfnamefont {L.}~\bibnamefont
  {Pollet}},\ }\bibfield  {title} {\bibinfo {title} {Diagrammatic monte carlo
  study of quasi-two-dimensional fermi polarons},\ }\href
  {https://doi.org/10.1103/PhysRevB.90.104510} {\bibfield  {journal} {\bibinfo
  {journal} {Phys. Rev. B}\ }\textbf {\bibinfo {volume} {90}},\ \bibinfo
  {pages} {104510} (\bibinfo {year} {2014})}\BibitemShut {NoStop}%
\bibitem [{\citenamefont {Cetina}\ \emph {et~al.}(2016)\citenamefont {Cetina},
  \citenamefont {Jag}, \citenamefont {Lous}, \citenamefont {Fritsche},
  \citenamefont {Walraven}, \citenamefont {Grimm}, \citenamefont {Levinsen},
  \citenamefont {Parish}, \citenamefont {Schmidt}, \citenamefont {Knap},\ and\
  \citenamefont {Demler}}]{cetina_16}%
  \BibitemOpen
  \bibfield  {author} {\bibinfo {author} {\bibfnamefont {M.}~\bibnamefont
  {Cetina}}, \bibinfo {author} {\bibfnamefont {M.}~\bibnamefont {Jag}},
  \bibinfo {author} {\bibfnamefont {R.~S.}\ \bibnamefont {Lous}}, \bibinfo
  {author} {\bibfnamefont {I.}~\bibnamefont {Fritsche}}, \bibinfo {author}
  {\bibfnamefont {J.~T.~M.}\ \bibnamefont {Walraven}}, \bibinfo {author}
  {\bibfnamefont {R.}~\bibnamefont {Grimm}}, \bibinfo {author} {\bibfnamefont
  {J.}~\bibnamefont {Levinsen}}, \bibinfo {author} {\bibfnamefont {M.~M.}\
  \bibnamefont {Parish}}, \bibinfo {author} {\bibfnamefont {R.}~\bibnamefont
  {Schmidt}}, \bibinfo {author} {\bibfnamefont {M.}~\bibnamefont {Knap}},\ and\
  \bibinfo {author} {\bibfnamefont {E.}~\bibnamefont {Demler}},\ }\bibfield
  {title} {\bibinfo {title} {Ultrafast many-body interferometry of impurities
  coupled to a fermi sea},\ }\href {https://doi.org/10.1126/science.aaf5134}
  {\bibfield  {journal} {\bibinfo  {journal} {Science}\ }\textbf {\bibinfo
  {volume} {354}},\ \bibinfo {pages} {96} (\bibinfo {year} {2016})}\BibitemShut
  {NoStop}%
\bibitem [{\citenamefont {Parish}(2011)}]{Parish2011}%
  \BibitemOpen
  \bibfield  {author} {\bibinfo {author} {\bibfnamefont {M.~M.}\ \bibnamefont
  {Parish}},\ }\bibfield  {title} {\bibinfo {title} {Polaron-molecule
  transitions in a two-dimensional fermi gas},\ }\href
  {https://doi.org/10.1103/physreva.83.051603} {\bibfield  {journal} {\bibinfo
  {journal} {Physical Review A}\ }\textbf {\bibinfo {volume} {83}},\ \bibinfo
  {pages} {051603} (\bibinfo {year} {2011})}\BibitemShut {NoStop}%
\bibitem [{\citenamefont {Parish}\ and\ \citenamefont
  {Levinsen}(2013)}]{Parish2013}%
  \BibitemOpen
  \bibfield  {author} {\bibinfo {author} {\bibfnamefont {M.~M.}\ \bibnamefont
  {Parish}}\ and\ \bibinfo {author} {\bibfnamefont {J.}~\bibnamefont
  {Levinsen}},\ }\bibfield  {title} {\bibinfo {title} {Highly polarized fermi
  gases in two dimensions},\ }\href
  {https://doi.org/10.1103/physreva.87.033616} {\bibfield  {journal} {\bibinfo
  {journal} {Physical Review A}\ }\textbf {\bibinfo {volume} {87}},\ \bibinfo
  {pages} {033616} (\bibinfo {year} {2013})}\BibitemShut {NoStop}%
\bibitem [{\citenamefont {Zöllner}\ \emph {et~al.}(2011)\citenamefont
  {Zöllner}, \citenamefont {Bruun},\ and\ \citenamefont
  {Pethick}}]{Zoellner2011}%
  \BibitemOpen
  \bibfield  {author} {\bibinfo {author} {\bibfnamefont {S.}~\bibnamefont
  {Zöllner}}, \bibinfo {author} {\bibfnamefont {G.~M.}\ \bibnamefont
  {Bruun}},\ and\ \bibinfo {author} {\bibfnamefont {C.~J.}\ \bibnamefont
  {Pethick}},\ }\bibfield  {title} {\bibinfo {title} {Polarons and molecules in
  a two-dimensional fermi gas},\ }\href
  {https://doi.org/10.1103/physreva.83.021603} {\bibfield  {journal} {\bibinfo
  {journal} {Physical Review A}\ }\textbf {\bibinfo {volume} {83}},\ \bibinfo
  {pages} {021603} (\bibinfo {year} {2011})}\BibitemShut {NoStop}%
\bibitem [{sup()}]{supp}%
  \BibitemOpen
  \href@noop {} {}\bibinfo {note} {See supplementary online
  material}\BibitemShut {NoStop}%
\bibitem [{\citenamefont {Sieberer}\ \emph {et~al.}(2016)\citenamefont
  {Sieberer}, \citenamefont {Buchhold},\ and\ \citenamefont
  {Diehl}}]{sieberer_16}%
  \BibitemOpen
  \bibfield  {author} {\bibinfo {author} {\bibfnamefont {L.~M.}\ \bibnamefont
  {Sieberer}}, \bibinfo {author} {\bibfnamefont {M.}~\bibnamefont {Buchhold}},\
  and\ \bibinfo {author} {\bibfnamefont {S.}~\bibnamefont {Diehl}},\ }\bibfield
   {title} {\bibinfo {title} {Keldysh field theory for driven open quantum
  systems},\ }\href {https://doi.org/10.1088/0034-4885/79/9/096001} {\bibfield
  {journal} {\bibinfo  {journal} {Reports on Progress in Physics}\ }\textbf
  {\bibinfo {volume} {79}},\ \bibinfo {pages} {096001} (\bibinfo {year}
  {2016})}\BibitemShut {NoStop}%
\bibitem [{\citenamefont {Wasak}\ \emph {et~al.}(2021)\citenamefont {Wasak},
  \citenamefont {Schmidt},\ and\ \citenamefont {Piazza}}]{wasak_21}%
  \BibitemOpen
  \bibfield  {author} {\bibinfo {author} {\bibfnamefont {T.}~\bibnamefont
  {Wasak}}, \bibinfo {author} {\bibfnamefont {R.}~\bibnamefont {Schmidt}},\
  and\ \bibinfo {author} {\bibfnamefont {F.}~\bibnamefont {Piazza}},\
  }\bibfield  {title} {\bibinfo {title} {Quantum-zeno fermi polaron in the
  strong dissipation limit},\ }\href
  {https://doi.org/10.1103/PhysRevResearch.3.013086} {\bibfield  {journal}
  {\bibinfo  {journal} {Phys. Rev. Research}\ }\textbf {\bibinfo {volume}
  {3}},\ \bibinfo {pages} {013086} (\bibinfo {year} {2021})}\BibitemShut
  {NoStop}%
\bibitem [{\citenamefont {Kohstall}\ \emph {et~al.}(2012)\citenamefont
  {Kohstall}, \citenamefont {Zaccanti}, \citenamefont {Jag}, \citenamefont
  {Trenkwalder}, \citenamefont {Massignan}, \citenamefont {Bruun},
  \citenamefont {Schreck},\ and\ \citenamefont {Grimm}}]{kohstall12}%
  \BibitemOpen
  \bibfield  {author} {\bibinfo {author} {\bibfnamefont {C.}~\bibnamefont
  {Kohstall}}, \bibinfo {author} {\bibfnamefont {M.}~\bibnamefont {Zaccanti}},
  \bibinfo {author} {\bibfnamefont {M.}~\bibnamefont {Jag}}, \bibinfo {author}
  {\bibfnamefont {A.}~\bibnamefont {Trenkwalder}}, \bibinfo {author}
  {\bibfnamefont {P.}~\bibnamefont {Massignan}}, \bibinfo {author}
  {\bibfnamefont {G.~M.}\ \bibnamefont {Bruun}}, \bibinfo {author}
  {\bibfnamefont {F.}~\bibnamefont {Schreck}},\ and\ \bibinfo {author}
  {\bibfnamefont {R.}~\bibnamefont {Grimm}},\ }\bibfield  {title} {\bibinfo
  {title} {Metastability and coherence of repulsive polarons in a strongly
  interacting fermi mixture},\ }\href {https://doi.org/10.1038/nature11065}
  {\bibfield  {journal} {\bibinfo  {journal} {Nature}\ }\textbf {\bibinfo
  {volume} {485}},\ \bibinfo {pages} {615} (\bibinfo {year}
  {2012})}\BibitemShut {NoStop}%
\bibitem [{\citenamefont {Massignan}(2012)}]{massignan12}%
  \BibitemOpen
  \bibfield  {author} {\bibinfo {author} {\bibfnamefont {P.}~\bibnamefont
  {Massignan}},\ }\bibfield  {title} {\bibinfo {title} {Polarons and dressed
  molecules near narrow feshbach resonances},\ }\href
  {https://doi.org/10.1209/0295-5075/98/10012} {\bibfield  {journal} {\bibinfo
  {journal} {{EPL} (Europhysics Letters)}\ }\textbf {\bibinfo {volume} {98}},\
  \bibinfo {pages} {10012} (\bibinfo {year} {2012})}\BibitemShut {NoStop}%
\bibitem [{\citenamefont {Takemura}\ \emph {et~al.}(2014)\citenamefont
  {Takemura}, \citenamefont {Trebaol}, \citenamefont {Wouters}, \citenamefont
  {Portella-Oberli},\ and\ \citenamefont {Deveaud}}]{takemura_14}%
  \BibitemOpen
  \bibfield  {author} {\bibinfo {author} {\bibfnamefont {N.}~\bibnamefont
  {Takemura}}, \bibinfo {author} {\bibfnamefont {S.}~\bibnamefont {Trebaol}},
  \bibinfo {author} {\bibfnamefont {M.}~\bibnamefont {Wouters}}, \bibinfo
  {author} {\bibfnamefont {M.~T.}\ \bibnamefont {Portella-Oberli}},\ and\
  \bibinfo {author} {\bibfnamefont {B.}~\bibnamefont {Deveaud}},\ }\bibfield
  {title} {\bibinfo {title} {Polaritonic feshbach resonance},\ }\href
  {https://doi.org/10.1038/nphys2999} {\bibfield  {journal} {\bibinfo
  {journal} {Nature Physics}\ }\textbf {\bibinfo {volume} {10}},\ \bibinfo
  {pages} {500} (\bibinfo {year} {2014})}\BibitemShut {NoStop}%
\bibitem [{\citenamefont {Laussy}\ \emph {et~al.}(2010)\citenamefont {Laussy},
  \citenamefont {Kavokin},\ and\ \citenamefont {Shelykh}}]{kavokin_10}%
  \BibitemOpen
  \bibfield  {author} {\bibinfo {author} {\bibfnamefont {F.~P.}\ \bibnamefont
  {Laussy}}, \bibinfo {author} {\bibfnamefont {A.~V.}\ \bibnamefont
  {Kavokin}},\ and\ \bibinfo {author} {\bibfnamefont {I.~A.}\ \bibnamefont
  {Shelykh}},\ }\bibfield  {title} {\bibinfo {title} {Exciton-polariton
  mediated superconductivity},\ }\href
  {https://doi.org/10.1103/PhysRevLett.104.106402} {\bibfield  {journal}
  {\bibinfo  {journal} {Phys. Rev. Lett.}\ }\textbf {\bibinfo {volume} {104}},\
  \bibinfo {pages} {106402} (\bibinfo {year} {2010})}\BibitemShut {NoStop}%
\bibitem [{\citenamefont {Cotle\ifmmode~\mbox{\c{t}}\else \c{t}\fi{}}\ \emph
  {et~al.}(2016)\citenamefont {Cotle\ifmmode~\mbox{\c{t}}\else \c{t}\fi{}},
  \citenamefont {Zeytino\ifmmode~\check{g}\else \v{g}\fi{}lu}, \citenamefont
  {Sigrist}, \citenamefont {Demler},\ and\ \citenamefont
  {Imamo\v{g}lu}}]{ovidiu_sc_16}%
  \BibitemOpen
  \bibfield  {author} {\bibinfo {author} {\bibfnamefont {O.}~\bibnamefont
  {Cotle\ifmmode~\mbox{\c{t}}\else \c{t}\fi{}}}, \bibinfo {author}
  {\bibfnamefont {S.}~\bibnamefont {Zeytino\ifmmode~\check{g}\else
  \v{g}\fi{}lu}}, \bibinfo {author} {\bibfnamefont {M.}~\bibnamefont
  {Sigrist}}, \bibinfo {author} {\bibfnamefont {E.}~\bibnamefont {Demler}},\
  and\ \bibinfo {author} {\bibfnamefont {A.}~\bibnamefont {Imamo\v{g}lu}},\
  }\bibfield  {title} {\bibinfo {title} {Superconductivity and other collective
  phenomena in a hybrid bose-fermi mixture formed by a polariton condensate and
  an electron system in two dimensions},\ }\href
  {https://doi.org/10.1103/PhysRevB.93.054510} {\bibfield  {journal} {\bibinfo
  {journal} {Phys. Rev. B}\ }\textbf {\bibinfo {volume} {93}},\ \bibinfo
  {pages} {054510} (\bibinfo {year} {2016})}\BibitemShut {NoStop}%
\bibitem [{\citenamefont {Wang}\ \emph {et~al.}(2019)\citenamefont {Wang},
  \citenamefont {Rhodes}, \citenamefont {Watanabe}, \citenamefont {Taniguchi},
  \citenamefont {Hone}, \citenamefont {Shan},\ and\ \citenamefont
  {Mak}}]{Wang19}%
  \BibitemOpen
  \bibfield  {author} {\bibinfo {author} {\bibfnamefont {Z.}~\bibnamefont
  {Wang}}, \bibinfo {author} {\bibfnamefont {D.~A.}\ \bibnamefont {Rhodes}},
  \bibinfo {author} {\bibfnamefont {K.}~\bibnamefont {Watanabe}}, \bibinfo
  {author} {\bibfnamefont {T.}~\bibnamefont {Taniguchi}}, \bibinfo {author}
  {\bibfnamefont {J.~C.}\ \bibnamefont {Hone}}, \bibinfo {author}
  {\bibfnamefont {J.}~\bibnamefont {Shan}},\ and\ \bibinfo {author}
  {\bibfnamefont {K.~F.}\ \bibnamefont {Mak}},\ }\bibfield  {title} {\bibinfo
  {title} {Evidence of high-temperature exciton condensation in two-dimensional
  atomic double layers},\ }\href {https://doi.org/10.1038/s41586-019-1591-7}
  {\bibfield  {journal} {\bibinfo  {journal} {Nature}\ }\textbf {\bibinfo
  {volume} {574}},\ \bibinfo {pages} {76} (\bibinfo {year} {2019})}\BibitemShut
  {NoStop}%
\bibitem [{\citenamefont {Efimov}(1970{\natexlab{a}})}]{efimov_res_70}%
  \BibitemOpen
  \bibfield  {author} {\bibinfo {author} {\bibfnamefont {V.}~\bibnamefont
  {Efimov}},\ }\bibfield  {title} {\bibinfo {title} {Energy levels arising from
  resonant two-body forces in a three-body system},\ }\href
  {https://doi.org/https://doi.org/10.1016/0370-2693(70)90349-7} {\bibfield
  {journal} {\bibinfo  {journal} {Physics Letters B}\ }\textbf {\bibinfo
  {volume} {33}},\ \bibinfo {pages} {563} (\bibinfo {year}
  {1970}{\natexlab{a}})}\BibitemShut {NoStop}%
\bibitem [{\citenamefont {Efimov}(1970{\natexlab{b}})}]{efimov_original_70}%
  \BibitemOpen
  \bibfield  {author} {\bibinfo {author} {\bibfnamefont {V.~N.}\ \bibnamefont
  {Efimov}},\ }\bibfield  {title} {\bibinfo {title} {Weakly bound states of
  three resonantly interacting particles},\ }\bibfield  {journal} {\bibinfo
  {journal} {Yadern. Fiz. 12: 1080-91(Nov 1970).}\ }\href
  {https://www.osti.gov/biblio/4068792} {} (\bibinfo {year}
  {1970}{\natexlab{b}})\BibitemShut {NoStop}%
\bibitem [{\citenamefont {Slagle}\ and\ \citenamefont {Fu}(2020)}]{slagle20}%
  \BibitemOpen
  \bibfield  {author} {\bibinfo {author} {\bibfnamefont {K.}~\bibnamefont
  {Slagle}}\ and\ \bibinfo {author} {\bibfnamefont {L.}~\bibnamefont {Fu}},\
  }\bibfield  {title} {\bibinfo {title} {Charge transfer excitations, pair
  density waves, and superconductivity in moir\'e materials},\ }\href
  {https://doi.org/10.1103/PhysRevB.102.235423} {\bibfield  {journal} {\bibinfo
   {journal} {Phys. Rev. B}\ }\textbf {\bibinfo {volume} {102}},\ \bibinfo
  {pages} {235423} (\bibinfo {year} {2020})}\BibitemShut {NoStop}%
\bibitem [{\citenamefont {Babadi}(2013)}]{Babadi2013PhD}%
  \BibitemOpen
  \bibfield  {author} {\bibinfo {author} {\bibfnamefont {M.}~\bibnamefont
  {Babadi}},\ }\href@noop {} {\emph {\bibinfo {title} {Non-equilibrium dynamics
  of artificial quantum matter}}}\ (\bibinfo  {publisher} {Harvard
  University},\ \bibinfo {year} {2013})\BibitemShut {NoStop}%
\end{thebibliography}
\end{document}